\documentclass[twocolumn]{aastex631}
\usepackage{natbib,graphics,graphicx,multirow,amsmath,cancel,color,xcolor,hyperref}
\usepackage{ulem}
\usepackage{bm}
\usepackage{float}
\usepackage{cleveref}
\crefname{equation}{Eq}{Eqs} 
\crefrangelabelformat{equation}{(#3#1#4--#5#2#6)}

\shorttitle{Acceleration of Solar Energetic Particles by the shock of Interplanetary Coronal Mass Ejection}

\begin{document}
\title{Acceleration of Solar Energetic Particles by the shock of Interplanetary Coronal Mass Ejection}

\correspondingauthor{Shanwlee Sow Mondal}
\email{shanwlee@prl.res.in\\
shanwlee.sowmondal@gmail.com}

\author[0000-0003-4225-8520]{Shanwlee Sow Mondal}
\affil{Astronomy and Astrophysics Division, Physical Research Laboratory, Ahmedabad 380009, India}
\affil{     Indian Institute of Technology, Gandhinagar, Gujarat 382355, India}

\author{Aveek Sarkar}
\affil{Astronomy and Astrophysics Division, Physical Research Laboratory, Ahmedabad 380009, India}

\author{Bhargav Vaidya}
\affil{Discipline of Astronomy, Astrophysics and Space Engineering, Indian Institute of Technology Indore, Simrol, Indore 453552, India}

\author{Andrea Mignone}
\affil{Dipartimento di Fisica, Universita di Torino, via P. Giuria 1, I-10125 Torino, Italy}

\begin{abstract}
 Interplanetary Coronal Mass Ejection (ICME) shocks are known to accelerate particles and contribute significantly to Solar Energetic Particle (SEP) events. We have performed Magnetohydrodynamic-Particle in Cell (MHD-PIC) simulations of ICME shocks to understand the acceleration mechanism. These shocks vary in Alfv\'enic Mach numbers as well as in magnetic field orientations (parallel \& quasi-perpendicular). We find that Diffusive Shock Acceleration (DSA) plays a significant role in accelerating particles in a parallel ICME shock. In contrast, Shock Drift Acceleration (SDA) plays a pivotal role in a quasi-perpendicular shock. High-Mach shocks are seen to accelerate particles more efficiently. Our simulations suggest that background turbulence and local particle velocity distribution around the shock can indirectly hint at the acceleration mechanism. Our results also point towards a few possible \textit{in situ} observations that could validate our understanding of the topic.

\end{abstract}
\section{Introduction}\label{sec:introduction}
Solar Energetic Particles (SEPs) are non-thermal particles emanating from the Sun. Their energy can be several orders of magnitude higher than the ambient solar wind particles. These particles can be electrons, protons, or heavy ions. Usually, SEPs are detected soon after solar flares or when the heliospheric particle detectors encounter Interplanetary Coronal Mass Ejections (ICMEs). They have been detected at different heliospheric distances by multiple \textit{in situ} spacecrafts. SEPs can achieve energies up to several hundred MeV \citep{Reames_1999} and sometimes even to GeV \citep{Ryan_2000}. Intense SEP events may become hazardous for humans in space and can also damage the electronics of space instruments \citep{Feynman_2000}. \\

These energetic events are considered to originate via two different mechanisms -- particle acceleration in CME-driven shock waves (Gradual SEP) and magnetic reconnection (Impulsive SEP) \citep{Reames_1999}. Impulsive SEP events occur for a shorter duration ($\leq$ one day), with an occurrence frequency of about 1000 per year during solar maxima \citep{Reames_2002}. These are of low intensities. Gradual SEP events continue for a longer duration (several days). They are relatively rare (a few tens per year)\citep{Reames_2002} and produce proton flux several orders of magnitude larger than impulsive SEP events \citep{Klein_2017}. It is not easy to distinguish between these two processes in practice as they are intertwined and may co-occur. Magnetic reconnection may happen at the current sheet located behind the CME and at regions where the CME interacts with the ambient solar wind or fragmented current sheets of the post-eruptive environment. Shock acceleration may occur in the region where the reconnection jet interacts with the ambient coronal material and produces an outflow termination shock and shock waves driven by the super magnetosonic motion of the CME~\citep{Klein_2017}. 

Multiple \textit{in situ} observations have demonstrated (e.g., \cite{Sheeley_1983}; \cite{Gopalswamy_2003}; \cite{Kahler_2004}) the existence of SEPs during the passage of ICMEs. It is believed that the solar wind particles interacting with the shocks ahead of the ICMEs may get energized to such high energies \citep{Lee_1982}. \\

SEPs' peak intensities are found to be correlated with the speed and other physical parameters of the associated CMEs \citep{Kouloumvakos_2019, Desai_2016,Kahler_2001,Kahler_2005,Kahler_2013}. The strength of ICME driven shock waves is usually determined by the Alfv\'enic Mach number ($M_A=u/v_A$), where $v_A = B/\sqrt{4 \pi \rho}$ is the local solar wind Alfv\'en speed, and $u$ is the upstream solar wind speed in the shock rest frame. Commonly observed ICME-driven shocks possess $M_A \approx$ 2-4 \citep{Berdichevsky_2000, Oh_2007}. However, for stronger ICMEs, $M_A$ can go beyond five \citep{Lugaz_15}. A powerful shock was recorded \citep{Russell_2013} on 2012 July 23, by the STEREO-A spacecraft, where the Alfv\'enic Mach number was determined to be $\approx$ 21 \citep{Riley_16}. Proton energy spectra obtained from the particle detectors onboard STEREO-A have shown that particles are energized up to 100 MeV \citep{Russell_2013}.\\

Like other astrophysical shocks, ICME shocks are also collisionless in nature. The particle mean-free path of the system is much larger than the shock length scale (say, the width of the shock) or gyroradii of the particles. Primarily two mechanisms are thought to be responsible for particle acceleration in these kinds of systems -- Diffusive Shock Acceleration (DSA) \citep{Fermi_1954, Krymskii_1977, Bell_1978, Blandford_1978} and Shock Drift Acceleration (SDA) \citep{Armstrong_1985, Decker_1988}. Depending on the upstream magnetic field's orientation with respect to the shock normal, the respective mechanism dominates. For example, in the case of quasi-parallel shocks, where the ambient magnetic field makes an angle in the range $0 ^{\circ} < \theta < 45 ^{\circ}$ with the shock normal, DSA works as the dominant mechanism in accelerating the charged particles. Particles get scattered by magnetic turbulence and undergo repetitive head-on interactions with upstream and downstream plasma across the shock, thereby gaining energy. DSA is considered to be the most efficient mechanism responsible for the origin of highest-energy particles,  mostly observed at quasi-parallel shocks. For strong ($M_A \gg 1$) non-relativistic quasi-parallel shocks, the energy spectra of DSA maintains a power-law, $f(E) \propto E^{-1.5}$ \citep{Caprioli_2014a}. DSA is considered to be the promising theory behind the genesis of gradual solar energetic particle (SEP) events. \\

On the other hand, in quasi-perpendicular shocks, where the angle between the upstream magnetic field and the shock normal lies in the range $45 ^{\circ} < \theta < 90 ^{\circ}$, SDA plays a pivotal role in particle acceleration. In this case particles gain energy by drifting along the shock in the direction of the induced electric field~\citep{Decker_1983, Reames_2012}.\\

Particle acceleration is common in other astrophysical shocks such as in supernova remnants \citep{2012SSRv..173..369H} or in extragalactic jets \citep{2017SSRv..207....5R}. These shocks are considered to be the primary sources of galactic cosmic rays (CRs). Many efforts have already been made to understand the origin and transport of high energy particles (CRs) — their complex non-linear interactions with the background thermal plasma and the ambient magnetic field \citep{Fermi_1949,Fermi_1954,Blandford_1978}. \\

The most obvious framework to study the evolution of CRs numerically around a shock would be Particle in Cell (PIC) method. In the PIC, electrons and ions are treated as particles, therefore suitable to study plasma kinetics in the presence of a background electromagnetic field \citep{Spitkovsky_2005}. However, one needs to resolve the electron skin depth in the PIC code, which makes the simultaneous study of shock evolution and particle acceleration highly expensive. \\

Recently the field has been enriched with the development of hybrid-PIC codes where the electrons are considered as part of the background thermal fluid, and ions are treated as kinetic particles \citep{Lipatov_2002, Gargate_2007}. Such simulations need to resolve the ion-skin depth of the medium, making the simulations computationally expensive. With the availability of powerful machines, numerous works have been done where simulations were performed \citep{Caprioli_2013, Caprioli_2014a, Caprioli_2014b, Caprioli_2014c} with relatively large domains and for long duration, but still, only the initial stages of shock evolution and particle acceleration have been captured. However, the transition of charged particles from non-relativistic to relativistic regimes and the fraction of ions participating in the process would only be known when we can study the long term evolution of collisionless shocks. The MHD-PIC approach is more relevant for such purposes \citep{Bai_2015, Vanmarle_2018, Mignone18}. This method can describe the non-linear interaction between thermal (ions+electrons) and non-thermal energetic particles. The method conveniently ignores microscopic plasma scales but captures the kinetic effect of ions by resolving the particles' gyroradii. \\

The MHD formulations carry out the evolution of the thermal plasma consisting of both electrons and ions, and the dynamics of the particles (non-thermal ions) are studied using the PIC method. All the non-thermal particles are extracted from the thermal fluid itself. After extracting a certain fraction of ions from the background plasma, the Lorentz force experienced by these charged particles in the background fluid's electric and magnetic fields is calculated. This force now acts as the feedback force that the particles exert on the background fluid and thus modifies its evolution. Recent usage of adaptive mesh refinement on the background thermal plasma \citep{Vanmarle_2018} has made the MHD-PIC more computationally affordable method. \\

Numerical study of particle acceleration in heliospheric shocks was initiated using the Monte-Carlo simulation \citep{Baring_1997}. They showed that the particle energy spectra obtained from their simulation are in  good agreement with the \textit{in situ} energy spectra obtained from the interplanetary shock, suggesting an active DSA mechanism in the quasi-parallel heliospheric shock. \\

 An extensive study of Particle acceleration by CME shocks has been done using the Particle Acceleration and Transport in Heliosphere (PATH) code \citep{Zank_2000}. This model includes spherically symmetric Solar wind, in which a CME-driven shock wave propagates.  Particles are injected at the shock using an injection model where the number of injected particles is a small fraction of the thermal Solar wind particles. Injected particles get accelerated through DSA. Some of these particles also escape far upstream through diffusion. The outcome of such modeling includes the temporal evolution of energetic particle spectra and intensity profile at all the spatial locations (both upstream and downstream), and the determination of the particle injection energy and the maximum energy of particles accelerated at the shock. \\

The same PATH code was later modified to include shocks with arbitrary strengths \citep{Rice_2003}, giving a better estimation of the maximum particle energy for each shock. The transport of energetic particles \citep{Li_2003} escaping the shock, simulating the acceleration of heavy ions \citep{Li_2005b} also have been studied using the same. It has also been used \citep{Zank_2004, Zank_2006} to study particle acceleration at a perpendicular interplanetary shock. Modeling individual SEP events using the PATH code have been pursued by several authors \citep{Li_2005a, Zank_2007, Verkhoglyadova_2009, Verkhoglyadova_2010}. \cite{Li_2012} have studied the particle acceleration in oblique shock. Their results suggest that close to the Sun ($r < $ 0.1 AU), quasi-parallel shocks are better particle accelerators than quasi-perpendicular shocks. \\

 A 2-D extension of the original PATH code \citep{Zank_2000} was developed by \citep{Hu_2017}. This new model can take care of the evolution of the background solar wind and propagate the CME shock in a 2-D domain. It also calculates the particle acceleration at the shock along with their diffusion and convection in the upstream and downstream regions. This 2-D model can also capture the longitudinal distribution of the CME. \\

   Using the hybrid-PIC code, dHybrid \cite{Gargate_2007} have studied the particle acceleration in ICME shocks. This kind of study can be categorized more as a local shock simulation. Instead of taking the whole longitudinal distribution of the shock, a particular region is considered for the simulation, and acceleration of non-thermal particles is studied. The simulation could successfully capture the detailed spatial and temporal information of the electromagnetic waves generated due to counter-streaming ions in the shock upstream. However, the hybrid method's high demand for computational resources restricts them to simulate only up to 15 sec of the particle acceleration. The MHD-PIC method is a natural choice to deal with such difficulties. \\

The present work utilizes the MHD-PIC version of the PLUTO code \citep{Mignone18}. Unlike the hybrid-PIC code, through this approach, one can track the shock for several minutes and evolve the energy spectrum of non-thermal particles.  We simulate ICME shocks with various strengths (with different $M_A$) and magnetic configurations (parallel shock as well as perpendicular shock). The evolution of non-thermal protons is studied until their energy spectrum gets saturated. The aforementioned simulations explain the physical processes behind gradual SEP events. \\

The rest of the article is organized as follows. We describe the overall simulation setup in section~\ref{sec:setup}. Parallel and quasi-perpendicular shocks with high Mach numbers are elaborated in section~\ref{sec:mach_19}. Results of low Mach shocks are detailed in section~\ref{sec:lowmach}. Finally, we summarize our results in section~\ref{sec:summary}.

\section{Simulation setup}\label{sec:setup}
A complete derivation of the MHD-PIC formalism is given in \cite{Bai_2015}. The MHD-PIC version of the PLUTO code is also explained in \cite{Mignone18}. \\

We have simulated ICME shocks of different Alfv\'enic Mach numbers and analysed particle acceleration in each of them. ICME-shocks observed at 1 AU vary over a wide range of Alfv\'enic Mach numbers. A comparative study is performed to understand the role of the shocks' strength  in the particle energisation process. In all our simulations density, temperature and magnetic field scales are taken to be $n_{0} = 10$ cm$^{-3}$, $T_{0} = 5.6\times 10^{4}$ K and $B_{0}=2.817$ nT. Our length, time and velocity scales are chosen in terms of ion skin depth ($c/\omega_{pi}=72$ km), inverse of ion cyclotron frequency ($\Omega ^{-1}=3.69$ sec) and Alfv\'en velocity ($v_{A}=19.5$ km sec$^{-1}$). Here, $c$ is the speed of light and $\omega_{pi}$ is the ion plasma frequency. To emulate the solar wind (SW) plasma at $1$ AU, simulation domains are initially filled with a plasma of uniform density $n_{bg}=0.5 n_0$ ($5$ cm$^{-3}$), and temperature $9 T_{0}$ ($5 \times 10^{5}$ K) with uniform magnetic field strength of $B_{bg}=3.55 B_0$ ($10$ nT) whose orientation is different for parallel and quasi-perpendicular cases.  Each simulation domain consists of a 2D rectangular box of length $L_{x} = 1.5 \times 10^{5}(c/\omega_{pi})$ and $L_{y} = 5 \times 10^{3}(c/\omega_{pi})$ with uniform resolution $\Delta x = \Delta y = 10 ( c/\omega_{pi})$.

\subsection{Shock generation}\label{sec:shock_generation}
Generally, in the inner heliosphere, ICMEs travel faster than the ambient solar wind, with both reaching supersonic and super Alfv\'enic speeds. A shock wave gets developed ahead of the ICME. In between the shock and the ICME, a turbulent region gets matured known as the sheath region \citep{Kilpua_17}. \\

In numerical simulations such as the ones considered in this work, shocks are typically formed by using the piston method \citep{Gargate_2014, Bai_2015}. A leftward super Alfv\'enic flow driven continuously from the right boundary gives rise to a rightward propagating shock after getting reflected from the left conductive wall. The shock's strength and speed are controlled by adjusting the velocity of the upstream flow. The top and bottom boundaries maintain the periodic boundary condition.\\

The method described above can simulate an ICME shock quite nicely. Using this method, we capture the shock and part of the sheath region of an ICME in the downstream reference frame.  With other parameters chosen to emulate solar wind parameters at 1 AU, the inflow velocity is identified with the relative velocity between the SW and the CME. For our purpose we have chosen shocks with strong compression ratio, $r=\frac{\rho_d}{\rho_u}\approx \frac{\gamma + 1}{\gamma -1} \approx 4$, where $\gamma=5/3$ is the  adiabatic index and $\rho_u$ \& $\rho_d$ represent the upstream and downstream densities, respectively.\\

The shock converts the kinetic energy of the upstream flow to the thermal energy of the downstream plasma. As a result the downstream temperature is increased to $T_{d} = v_{u}^{2}/(r-1) \sim v_{u}^{2}/3$ (in code units), where $v_u$ is the upstream velocity in the downstream reference frame. For a strong shock the upstream temperature ($T_{u}$) is independent of the downstream temperature~\citep{Bai_2015}.

\subsection{Particle Injection}\label{sec:injection}

Unlike the Hybrid-PIC, the present MHD-PIC is not capable of generating non-thermal particles self-consistently from the thermal plasma. Therefore a strategy has to be adopted to inject protons that participate in the acceleration process. Even though electrons do also get energized in reality, limited computational power prohibits one from simulating their dynamics. Therefore we assume all electrons are thermalized and do not participate at the energization process. Only protons are capable of participating in the non-thermal process. In this article, proton and particle both represent the same entity. The participating protons are injected throughout the simulation. Equivalent mass, momentum, and energy of protons are subtracted from the background to ensure conservation. \\

The amount of injected protons is chosen in accordance with \cite{Bai_2015}, inspired by \cite{Caprioli_2014a}. The fraction of the injected proton mass to the mass swept away by the shock, $\eta$ is chosen to be $3 \times 10^{-3}$ in all our production runs. We have checked that if we increase this fraction by a significant amount, the injected protons can disrupt the shock through backreaction. On the other hand, a too-small value of $\eta$ may not contribute to the backreaction, and the proton acceleration process may become inefficient. The location of the shock front is tracked every time using the PLUTO shock tracking algorithm \citep{Mignone18}. Once the shock is detected, the amount of particles we inject is a fixed fraction, $\eta$ of the mass swept by the shock as mentioned above. The number of injected particles is equivalent to 10 particles per cell at background fluid density $n_{0}$. \\

Since the background thermal plasma and injected particles are treated differently in the MHD-PIC code, an ad hoc injection prescription needs to be adopted for proton injection. An injection recipe can be adopted by calibrating the MHD-PIC simulation with a more self-consistent hybrid-PIC simulation. Nevertheless, in the absence of such results, so far monoenergetic injection recipes have been used in MHD-PIC \citep{Bai_2015, Mignone18, Vanmarle_2018}. However, to inject particles with a wide range of energies, we have primarily used a Maxwellian distribution with downstream characteristic temperature. \\

The following sections are dedicated to the numerical setup and results obtained from simulations of particle acceleration in different shocks. \\

\section{Simulation of ICME shock with $M_{A} \approx 19 $}\label{sec:mach_19}
\subsection{Numerical setup} \label{sec:set_up}
Keeping in mind the detection of a strong ICME shock \citep{Russell_2013, Riley_16} at $1$ AU with Alfv\'enic Mach $\approx 19$ in the downstream reference frame, we initially simulate an ICME-shock with the same Alfv\'enic Mach number.  
    An initial background magnetic field of strength $B_{bg}=3.55 B_{0}$ is applied along the $x$ direction, which is parallel to the shock normal. Number of particles injected at the shock front depends on $\eta$ times the amount of mass swept by the shock, with $\eta = 3 \times 10^{-3}$ as mentioned in section \ref{sec:injection}. In a system like this, an instability is expected to develop due to the current generated by the energetic protons escaping the shock front thereby perturbing the upstream magnetic field \citep{Bell_2004}. The present resolution is sufficient to resolve the most unstable mode of such instability. Near the shock front the wavelength of this mode is approximately  $3\pi (B_{bg}/B_{0})/ \eta M_{A} = 587(c/\omega_{pi})$ \citep{Bai_2015}. Considering the SW and ICME speed to be $400$ km s$^{-1}$ and $2250$ km s$^{-1}$, respectively, we take their relative speed to be $-1850$ km s$^{-1}$.  Inflow speed of $-94.87 v_A$ ($= -1850$ km s$^{-1}$) produces a rightward propagating shock of Alfv\'enic Mach number $\approx 19$. The detailed generation mechanism of the shock is described in section \ref{sec:shock_generation}. Once the shock is formed,  particles are injected following the prescription discussed in section \ref{sec:injection}.\\

\begin{figure*}
\includegraphics[width=\textwidth,height=12cm,angle=0]{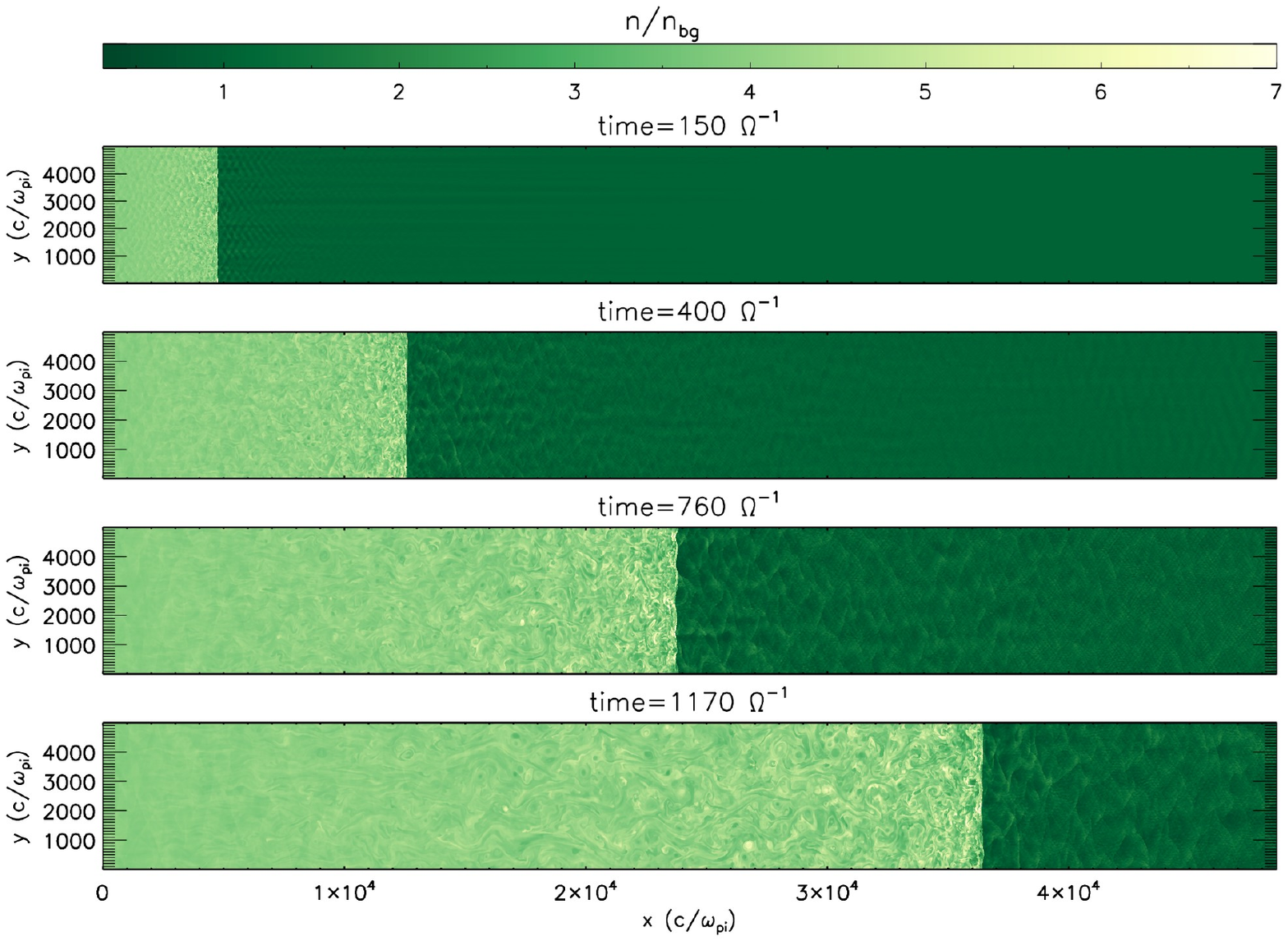}
\caption{Density evolution of the parallel shock ($M_A\approx19$), moving towards right. Only a fraction of the simulation domain is depicted here. (\href{https://www.youtube.com/watch?v=r9PHFcepczo}{An animation of this figure is available here}.)}
 \label{dens_snap}
\end{figure*}
 
    \begin{figure}
 \centering
    \includegraphics[width=0.48\textwidth,angle=0]{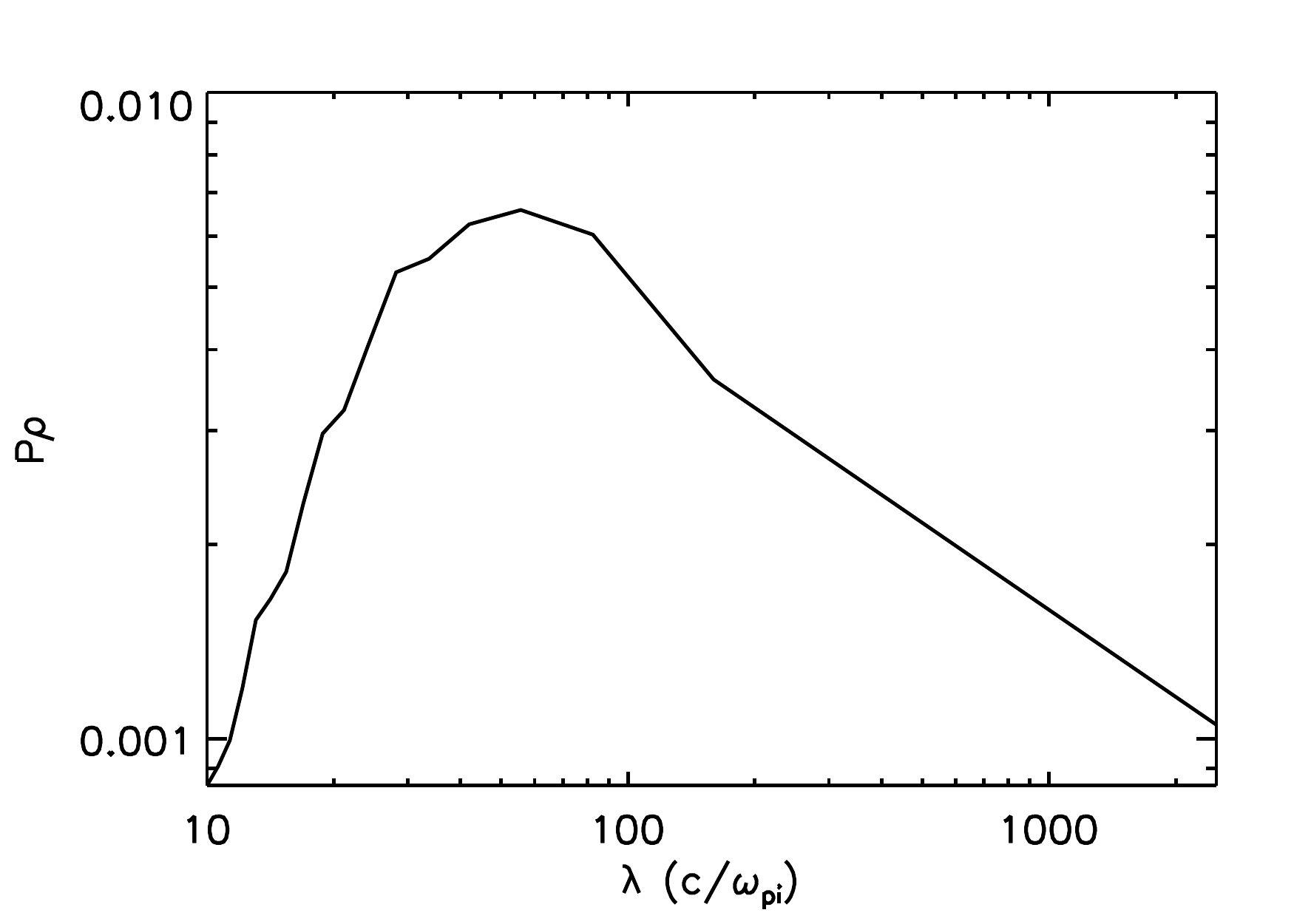}
  \caption{Fluctuating mass density ($\delta \rho/ \rho_{0}$) power spectrum during the near-saturation phase of the simulation (averaged over time $t=1000$ to $1170$ $\Omega^{-1}$ ). Here $\rho_{0}$ is the initial upstream density. The spectrum is derived at the location $330(c/\omega_{pi})$ ahead of the shock front (Figure~\ref{dens_snap}), over a width $2400( c/\omega_{pi})$.}
  \label{power_rho_norm}
  \end{figure}

\begin{figure*}
\includegraphics[width=\textwidth,height=12cm,angle=0]{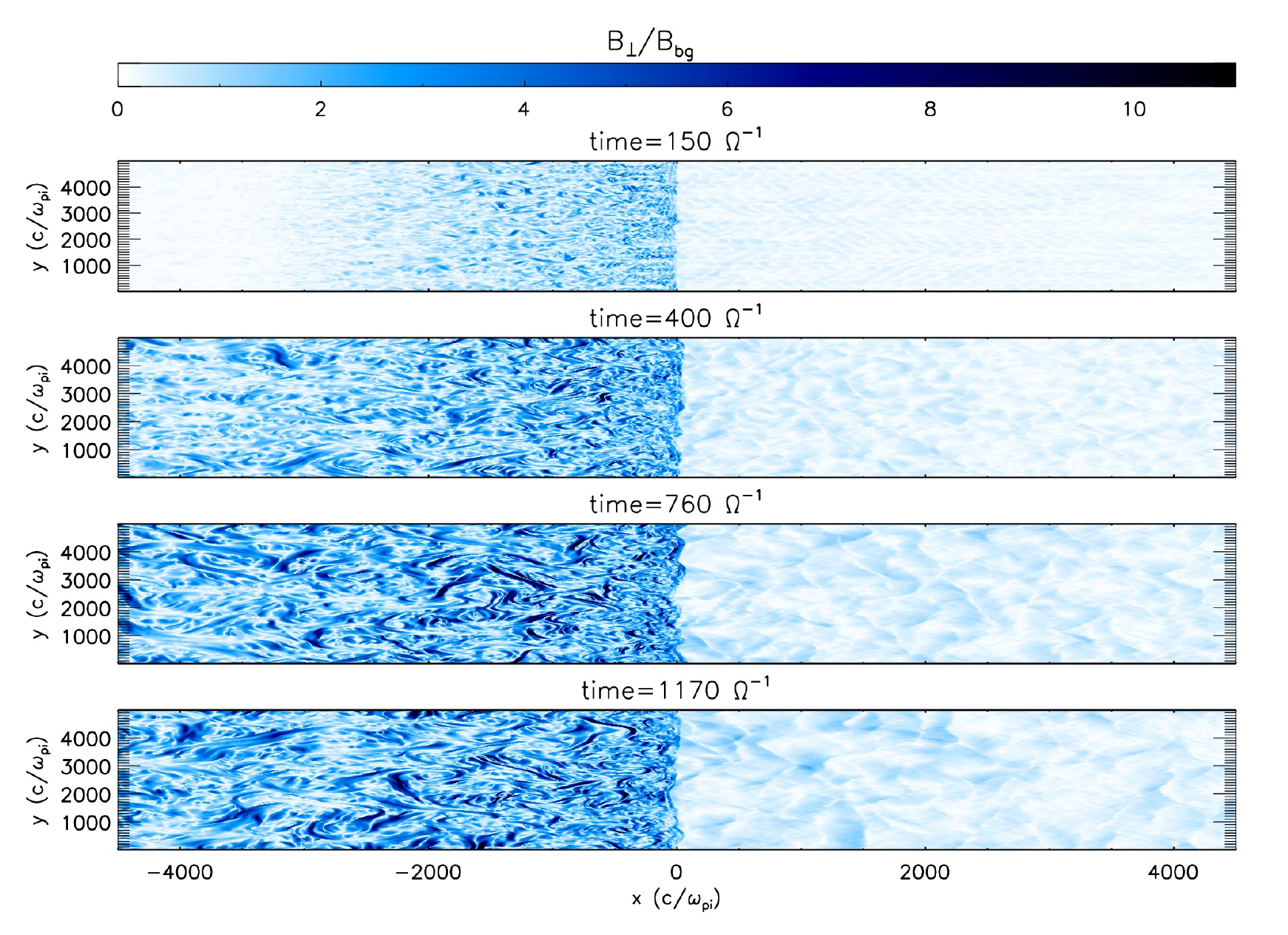}
\caption{Transverse magnetic field evolution around the shock front. At every time snap, the shock location is re-centered at $x=0$. (\href{https://www.youtube.com/watch?v=czJDW2NltpU}{An animation of this figure is available here.})}
\label{B_dens}
\end{figure*}

\begin{figure}
 \centering
    \includegraphics[width=0.48\textwidth,angle=0]{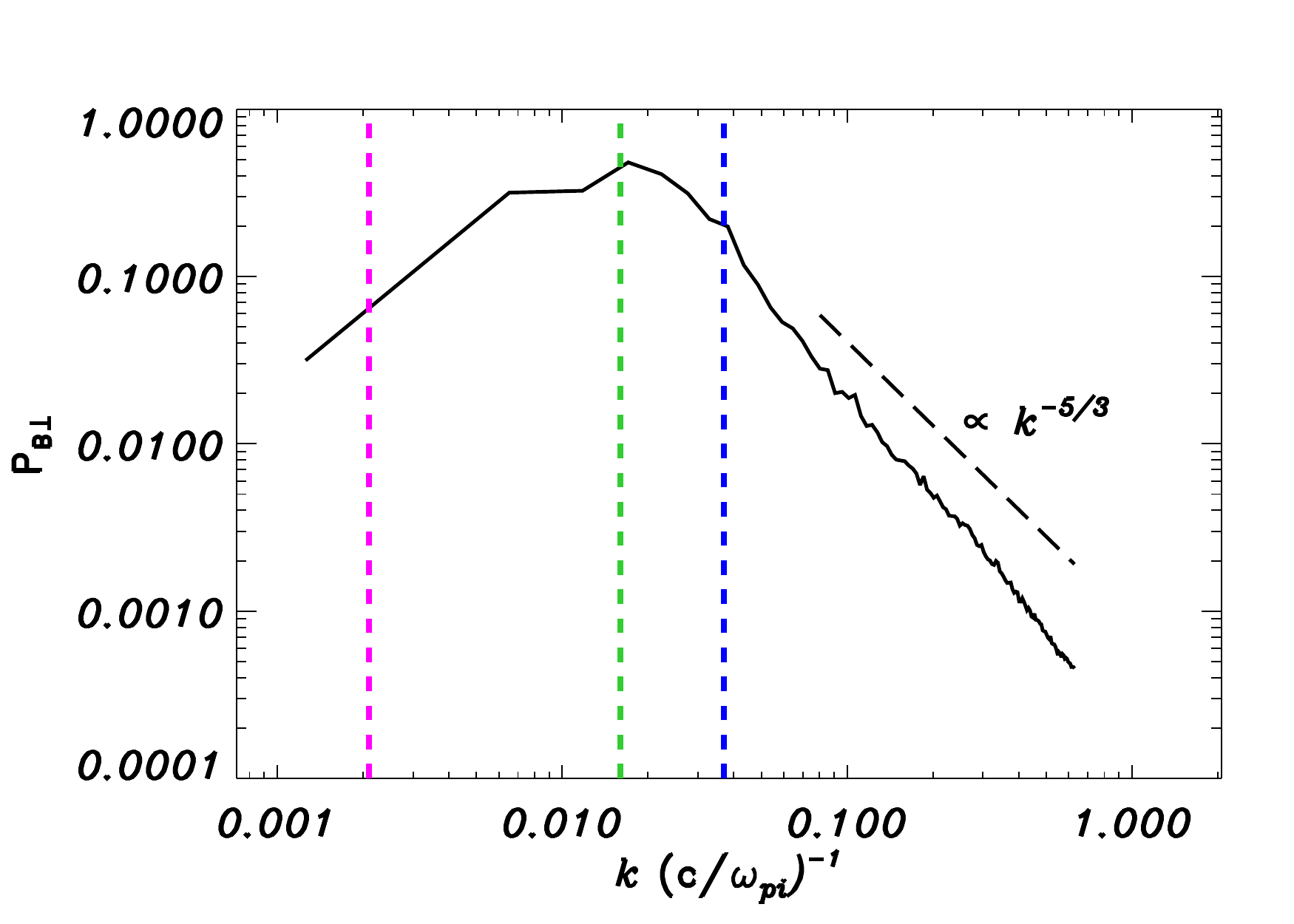}
  \caption{Dimensionless transverse magnetic energy power spectrum at $t=1170$ $\Omega^{-1}$ is calculated in a region of width $\approx 2400(c/\omega_{pi})$ just ahead of the shock. Blue and magenta dashed lines indicate  the wave modes in resonance with the particles having energies $E_{sh}$ and $300E_{sh}$ respectively. Theoretical prediction of the fastest growing mode of the Bell's instability is marked with the green dashed line, calculated with the transversely averaged current at the shock front. The black dashed line indicates the Kolmogorov power-law slope.}
    \label{bell}
  \end{figure}

\subsection{Shock propagation}

Density snapshots of the propagating shock are shown in Figure \ref{dens_snap}. It is to be noted that only a small part of the whole simulated domain is shown here. In the beginning, injected protons move freely in the upstream region. The motion of the energetic particles across the shock generates a current $\bm{J}$, primarily parallel to the ambient upstream magnetic field. The current initially perturbs the mean upstream magnetic field to generate Alfv\'en waves. Once the current gets stronger due to the high flux of energetic particles, it generates an instability known as Bell instability after its founder \citep{Bell_2004}. That, in turn, produces a perpendicular ($\delta \bm{B} \sim B_{\perp}$) magnetic field in the upstream \citep{Bell_2005, Reville_2012}. This magnetic field generates a force ($-\bm{J} \times \delta \bm{B}$), which acts on the local plasma and evacuates the region. As a result, low-density cavities are produced. Suprathermal particles occupy these cavities. Once the maximum current carrying particles' gyroradii become comparable to the cavities' size, the growth of the instability ceases \citep{Caprioli_2013}. Freely moving protons in the region get scattered by the density cavities.\\

To estimate the density cavities' size we consider the average density power spectrum over a short time window ($t=1000$ to $1170~\Omega^{-1}$), towards the end of the simulation. The density power spectrum (Figure~\ref{power_rho_norm}) of the shock upstream plasma is derived from a region $330(c/\omega_{pi})$ ahead of the shock front, having width $2400(c/\omega_{pi})$. Sufficient cavities have formed at this location. The peak of the power spectrum, which is at $55(c/\omega_{pi})$ indicates the dominant length scale of the cavities. The magnetic field at the same location averaged over the horizontal and transverse directions is found to be $\approx 3.97B_{0}$. Considering the typical particle velocity to be $\approx 232v_{A}$ (local particle velocity distribution peaks around this velocity), we find the particle-gyroradius to be $\approx 58 (c/\omega_{pi})$. The similarity in gyroradius and cavity size indicates the instability is saturated.\\ 

\begin{figure}
 \centering
    \includegraphics[width=0.48\textwidth,angle=0]{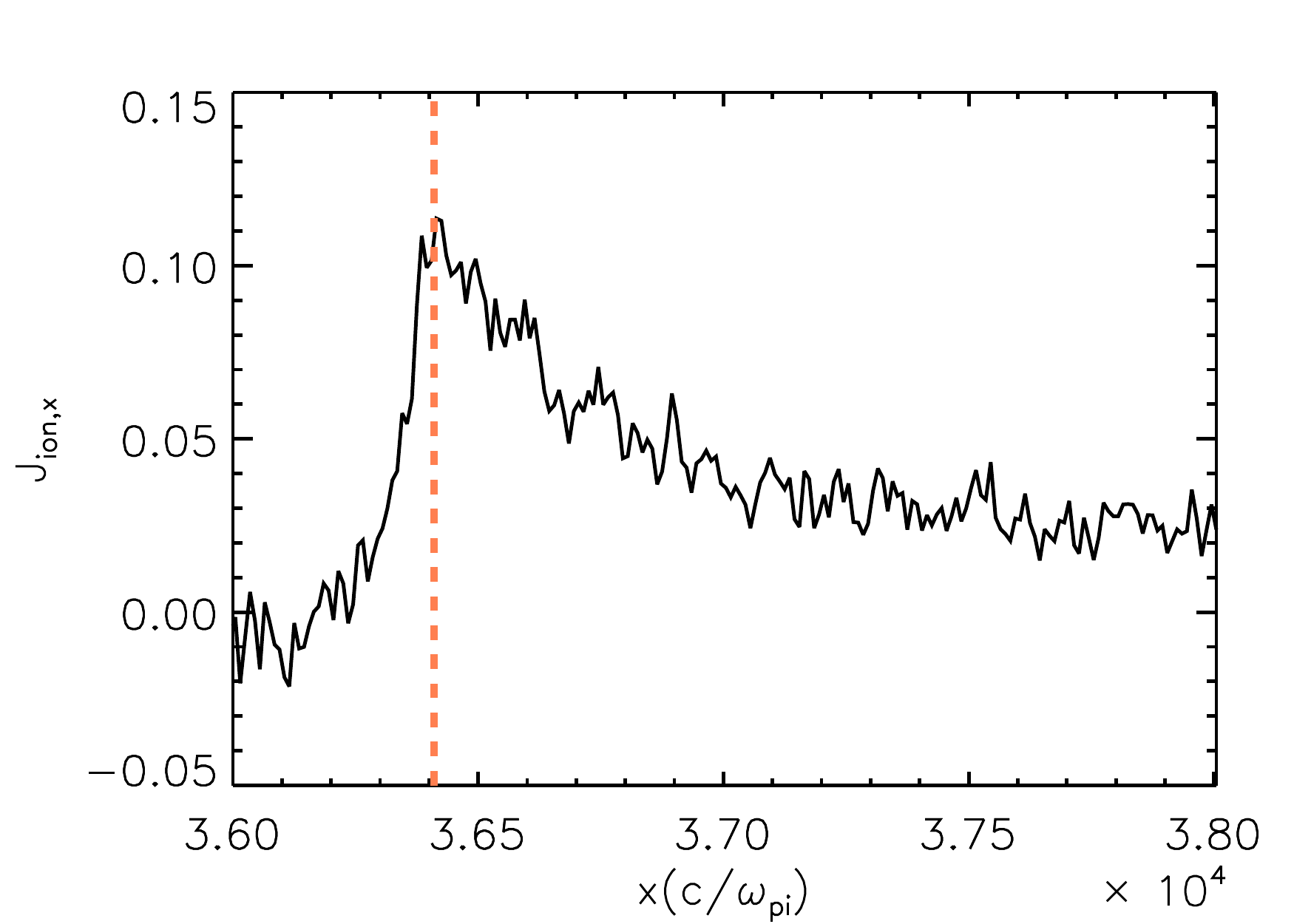}
  \caption{Transversely averaged $x$-component of the current density ($J_{ion,x}$) in normalized unit. The dashed line indicates the location of the shock and corresponding current is considered to calculate the theoretical value of the most unstable Bell mode (Figure~\ref{bell}).}
  \label{current}
  \end{figure}
\subsubsection{Role of instabilities in shock upstream}

Figure~\ref{B_dens} shows the magnetic field evolution close to the shock front. For every time snap, the shock front is re-centered at $x=0$. It is seen that in the shock precursor, along with the density, the magnetic field also becomes turbulent. The power spectrum of the perpendicular magnetic field, depicted in Figure~\ref{bell}, describes the nature of the magnetic turbulence. The maximum wave number ($k_{max}$) is determined by the simulation resolution, whereas the minimum wave number ($k_{min}$) is set by the largest length scale of the chosen region. Alfv\'en waves, initiated by the energetic particles in the shock upstream is natural to expect~\citep{Bell_2004}. 
These waves, resonating with the gyrofrequency of the streaming energetic particles, generate an instability in the system known as resonant streaming instability~\citep{Bell_2004,Amato_2009}. 
The mode of such instability achieves maximum growth rate at a wavenumber $k$ satisfying $kr_g (E, B_{bg})=1$, where $E$ is the energy of the particle gyrating with gyroradius $r_{g}$ in the background magnetic field of strength $B_{bg}$~\citep{Caprioli_2014b}.
We find that in the region the longest (for particles with $E=E_{sh}$) and shortest (particles with $E=300 E_{sh}$) wavenumbers of the resonant mode to be $3.74\times10^{-2}(\frac{c}{\omega_{pi}})^{-1}$ and $2.13\times10^{-3}(\frac{c}{\omega_{pi}})^{-1}$, respectively. While these two modes encompass the peak of the power spectrum (Figure~\ref{bell}), wave mode corresponding to the spectral peak resonates with the gyrofrequency of the particles having energy $E=3 E_{sh}$. However, the non-resonant hybrid (NRH) instability introduced by \cite{Bell_2004} also may exist in this atmosphere. The theoretical estimation of the fastest growing mode of this instability is found to be $k_{Bell}=\frac{J_{ion,x}}{2 B_{bg}}$, where $J_{ion,x}$ corresponds to the transversely averaged horizontal component of the current, generated due to the motion of the non-thermal protons, shown in Figure~\ref{current}. Considering the upstream shock vicinity value of $J_{ion,x}$ and the initial background  magnetic field $B_{bg}$, we estimate the fastest growing mode of the Bell instability to be $k_{Bell}= 1.59\times10^{-2}(\frac{c}{\omega_{pi}})^{-1}$, which nearly coincides with the wave mode corresponding to the peak of the transverse magnetic power spectrum. Among the two modes the non-resonant mode dominates when $ (\dfrac{J_{ion,x}r_{g}}{B_{bg}}) \gg 1$~\citep{Bell_2004, Amato_2009}.  Using the local values of $J_{ion,x}$, $r_g$ and $B_{bg}$ of the immediate shock upstream region, one finds the value of this non-dimensional quantity to be $\approx 1.5$, which favors the NRH instability in the region. Following the peak, the magnetic field power spectrum falls off with the characteristic Kolmogorov slope ($k^{-5/3}$).\\

\begin{figure}
  \centering
     \includegraphics[width=0.48\textwidth,angle=0]{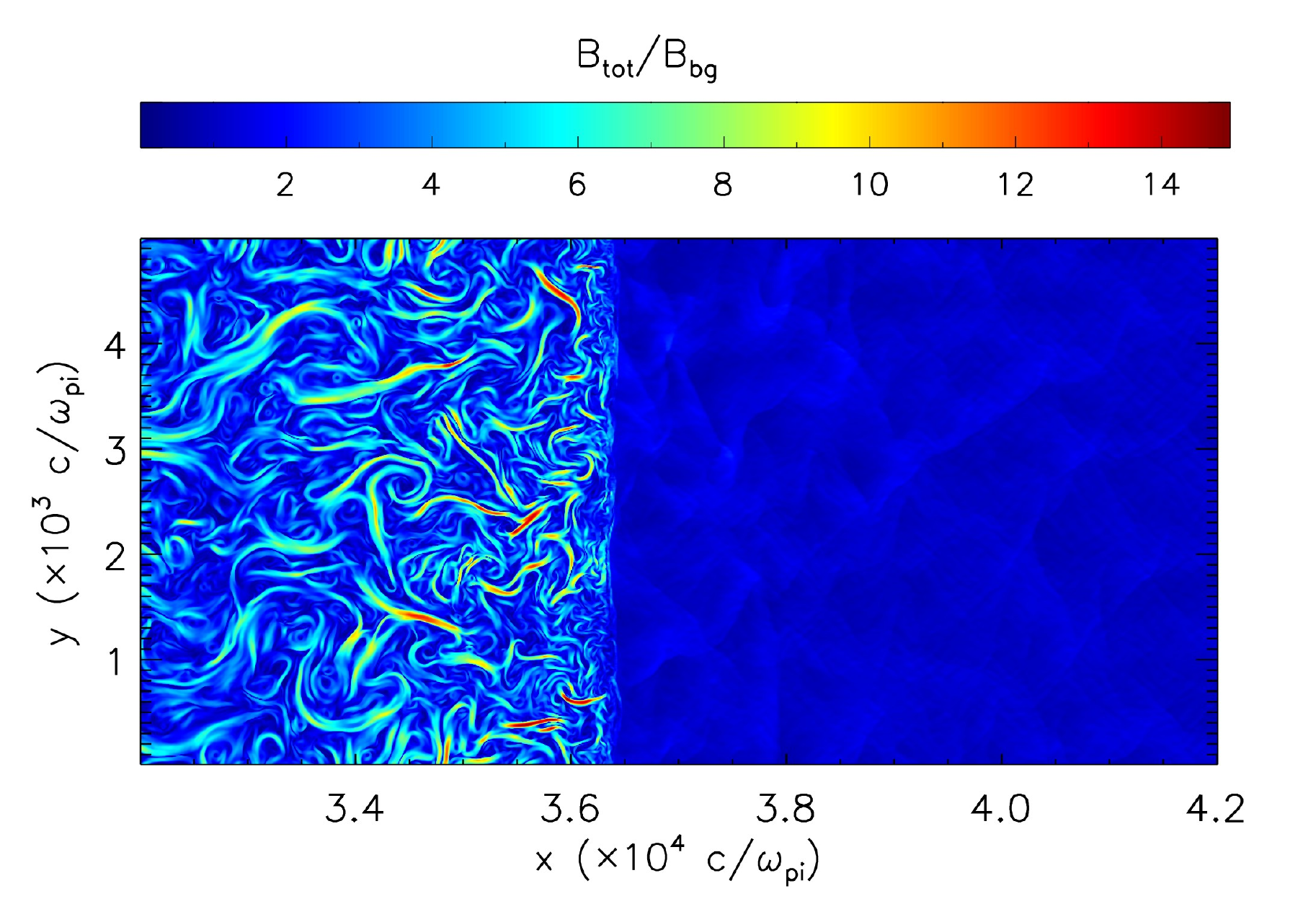}
  \caption{Total magnetic field contour close to the shock front at $t=1170~ \Omega^{-1}$}.
  \label{tot_B}
\end{figure}

Figure~\ref{tot_B} shows the magnetic field map in a region close to the shock front. Instabilities described above generate magnetic fluctuations enhancing the local magnetic field by as much as twice the initial value at the shock precursor. Following Rankine–Hugoniot shock condition, it is expected that the transverse magnetic components would get further amplified by a factor of 4. In other words, the already pre-amplified magnetic field of the upstream would be amplified further up to 8 times with respect to the background due to shock compression, and propagate into the downstream region. However, the total magnetic field contour in Figure~\ref{tot_B} shows an amplification by a factor of up to 15 at some locations of downstream. Clearly, a different magnetic field amplification mechanism is at play here. The initial shock surface gets perturbed by the upstream density disturbance and gets corrugated. At the corrugated surface, because of the non-aligned pressure and density gradient, vorticity gets introduced and drives the Richtmeyer Mechkov Instability (RMI)~\citep{Brouillette_02}. The turbulence introduced by the RMI helps to enhance the magnetic field further by stretching the field lines~\citep{Sano_12}, indicating the possible activation of turbulent dynamo behind the shock. The role of turbulent dynamo in the shock downstream has also been reported by earlier authors (e.g., \cite{Giacalone_07, Mizuno_11}).\\

\subsubsection{Magnetic field enhancement in the shock downstream}
\begin{figure}
 \centering
      \includegraphics[width=0.48\textwidth,angle=0]{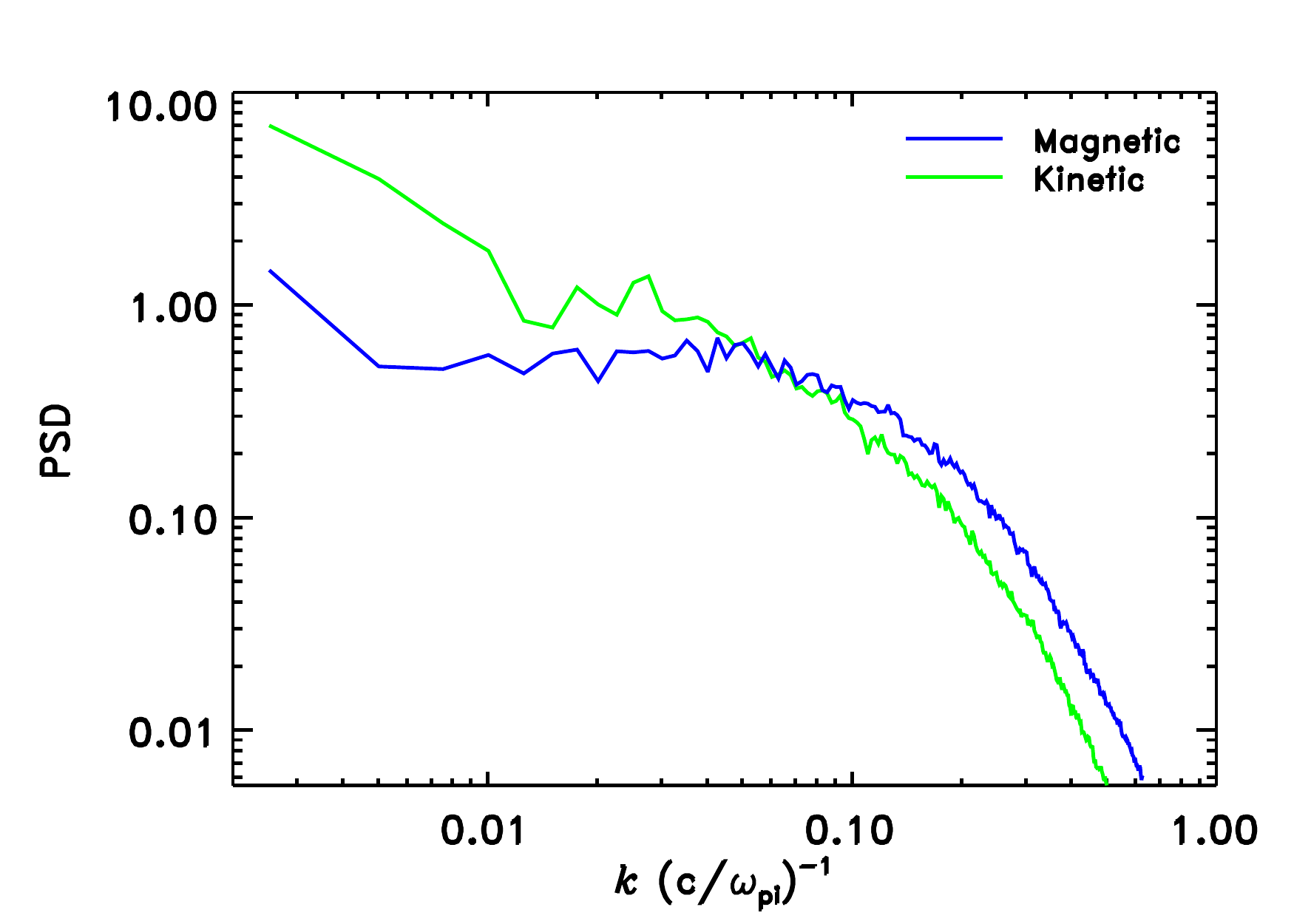}
  \caption{Dimensionless kinetic and magnetic energy power spectrum at $t = 1170~\Omega^{-1}$ in a region of width $8000(c/\omega_{pi})$ in the downstream region $400(c/\omega_{pi})$ away from the shock front. }
  \label{v_b}
  \end{figure}

For further investigation in the turbulence saturation phase, the kinetic and magnetic energy power spectrum of the region is plotted in Figure~\ref{v_b}. In the downstream region, $388(c/\omega_{pi})$ away from the shock front, a strip of width  $800(c/\omega_{pi})$ is considered to calculate the power spectrum. For the kinetic energy spectrum the parameter $\rho^{1/2}[\bm{u}-\bar{\bm{u}}]$ is employed \citep{Podesta_2007}, whereas for the magnetic energy spectrum $\bm{B}-\bm{\bar{B}}$ is considered. Here $\rho$, $\bm{u}$, $\bm{B}$, $\bar{\bm{u}}$ and $\bar{\bm{B}}$ are mass density, velocity, magnetic field, mean velocity and mean magnetic field of the region respectively. To calculate the mean fields ($\bar{\bm{u}}$,$\bar{\bm{B}}$), a time window of about $347~\Omega^{-1}$ is considered.  Higher magnetic energy towards the small scale is an indicator of possible small scale active dynamo in the region. \\

\subsubsection{Particle energization}

\begin{figure}
  \centering
     \includegraphics[width=0.48\textwidth,angle=0]{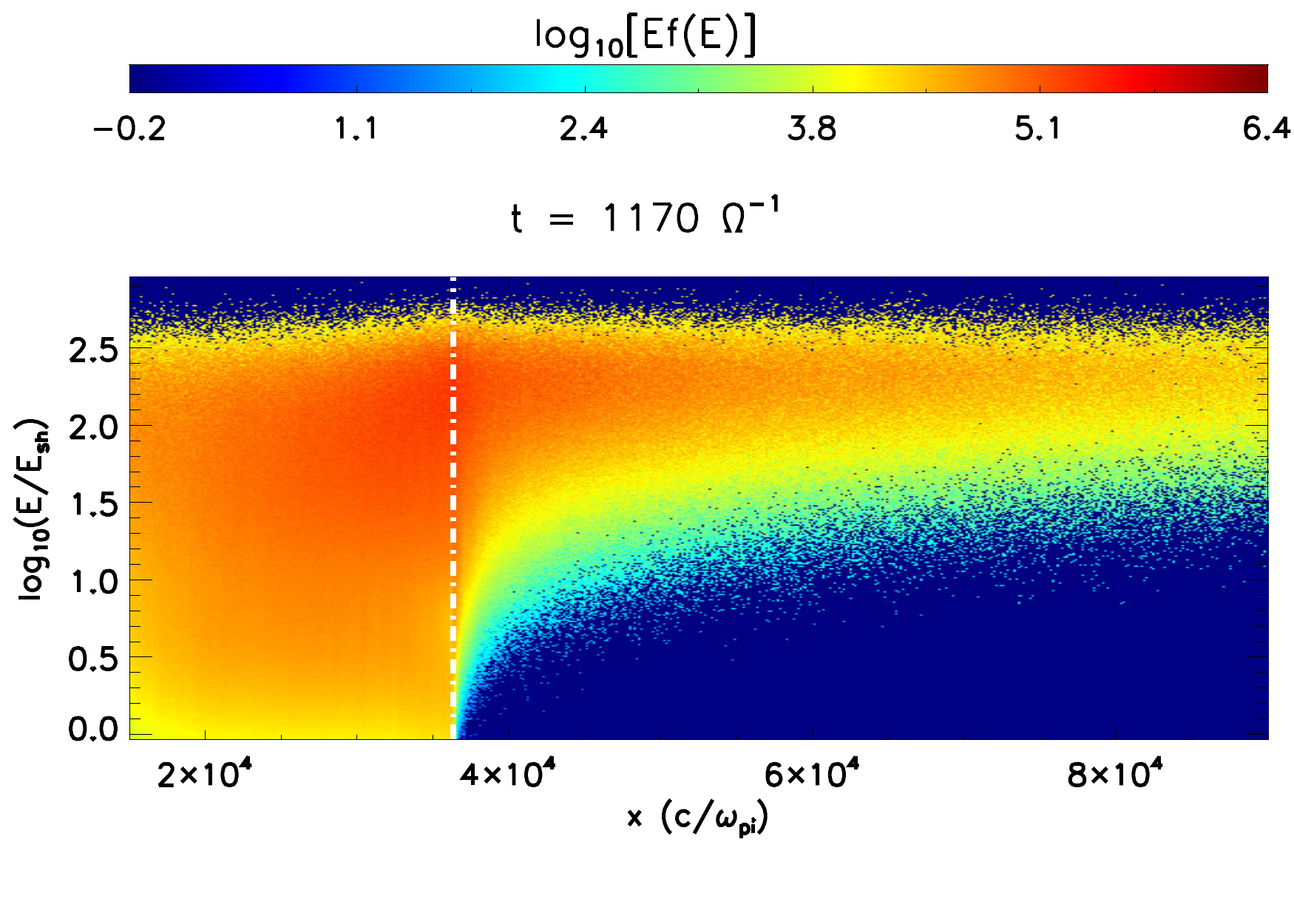}
  \caption{2D energy spectrum, showing the particle energy distribution as a function of position $x$. Here $E_{sh}$ is the shock energy given by $E_{sh} = v_{u}^2/2$ where $v_u$ is the upstream plasma speed in the downstream reference frame. Location of the shock front is indicated by the white dashed line. High energy particles outrunning the shock are evident from this 2D spectrum confirming the commencement of particle acceleration.}
  \label{2D_spectra}
  \end{figure}

\begin{figure}
  \centering
    \includegraphics[width=0.48\textwidth,angle=0]{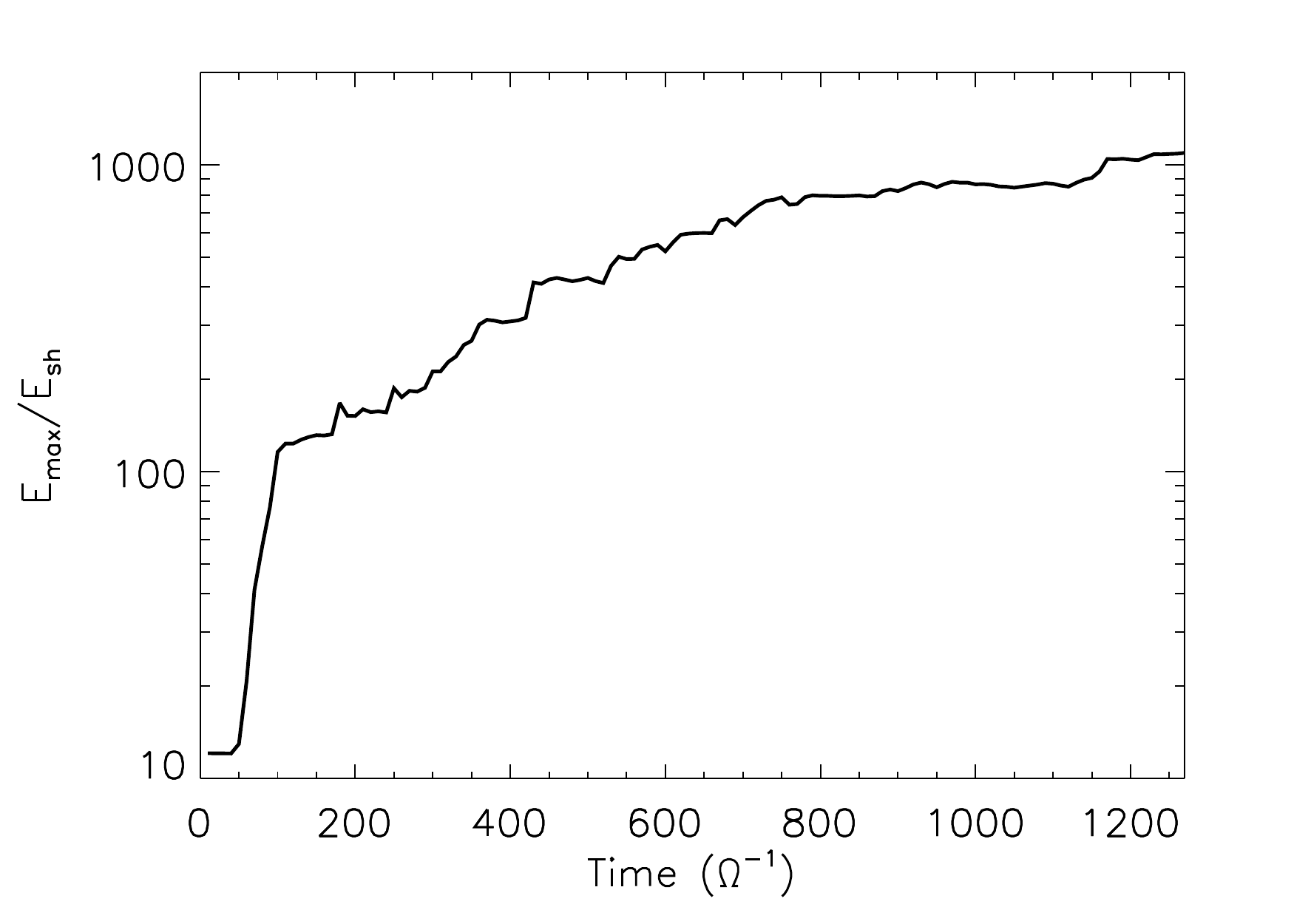}
   \caption{Time evolution of the maximum energy of the particles (normalized with respect to $E_{sh}$) present in the entire domain. The curve plateaus at large times suggesting near-saturation.}
  \label{e_max}
  \end{figure}

\begin{figure}
  \centering
    \includegraphics[width=0.48\textwidth,angle=0]{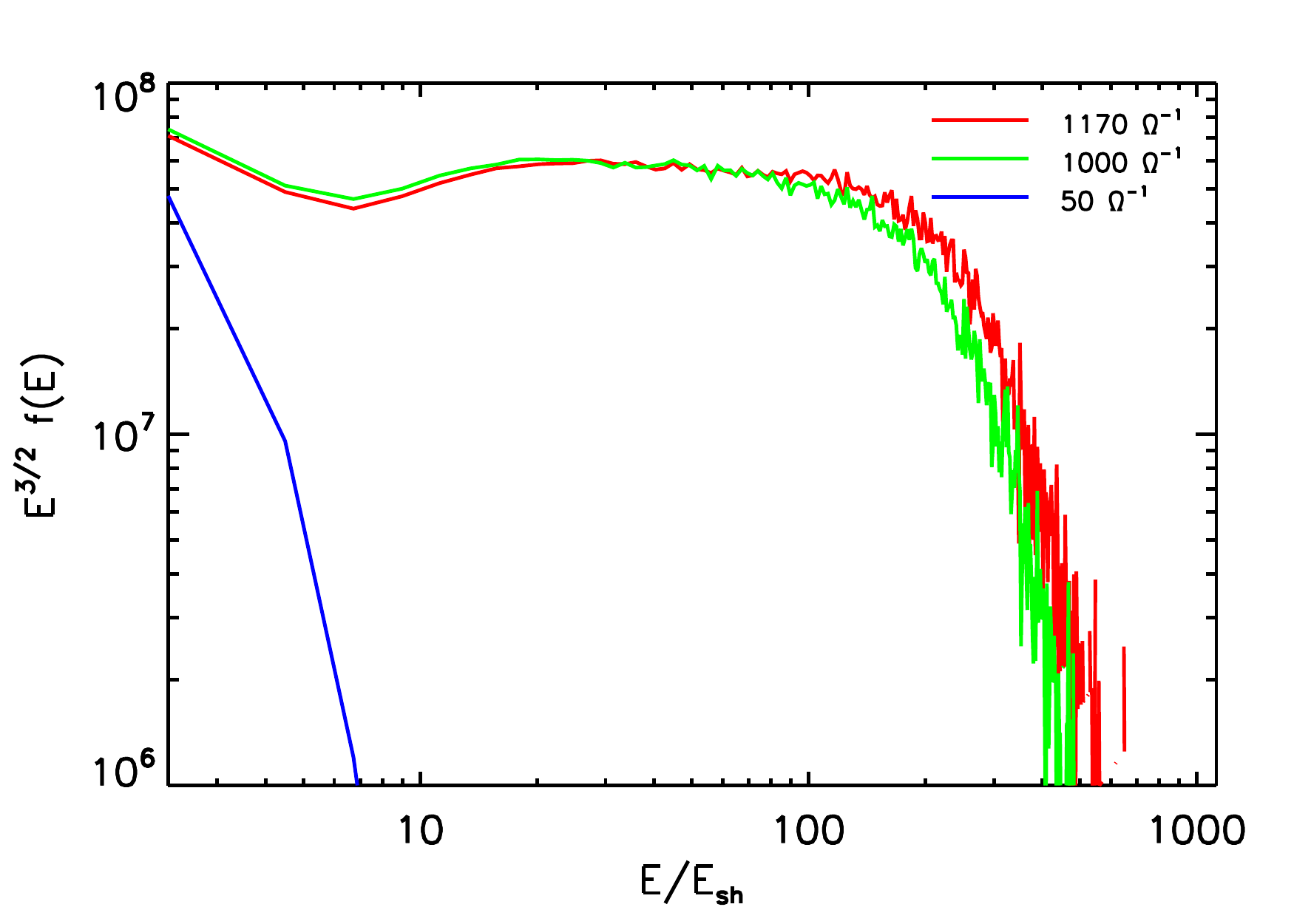}
  \caption{Evolution of downstream particle-energy spectra ($E^{3/2}$ compensated), extracted from a region of width $1000(c/\omega_{pi})$, just behind the shock. Flattening of the spectra ensures DSA is an efficient mechanism in particle acceleration. }
  \label{1D_spectra_par}
  \end{figure}

Turbulence in the shock upstream and downstream can scatter particles across the shock and energize them. As mentioned earlier, we inject Maxwellian distribution of particles with characteristic downstream temperature. A 2D particle energy spectrum at the evolved state of the simulation is shown in Figure~\ref{2D_spectra}. This figure depicts the particles' energy distribution around the shock. It clearly shows the existence of escaped high-energy particles in the shock upstream. \\

Figure~\ref{e_max} displays the maximum energy attained by the protons over time. We ran our simulation until maximum energy curve reaches its plateau. Figure~\ref{1D_spectra_par} depicts the evolution of 1D energy spectrum ($E^{3/2}$ compensated) of the particles. These spectra are extracted from a region of width $1000(c/\omega_{pi})$ just behind the shock. During the initial phase (e.g., $50~\Omega^{-1}$), the spectrum almost follows the injected Maxwellian distribution, eventually particles gain energy over time. During the near-saturation phase, particles are seen to have gained energy up to $E=500 E_{sh}$ ($\approx 10$ MeV) in the downstream region just behind the shock. The major part of the evolved spectrum follows $-3/2$ slope, indicating the active presence of the DSA mechanism. \\

\begin{figure}
  \centering
     \includegraphics[width=0.48\textwidth,angle=0]{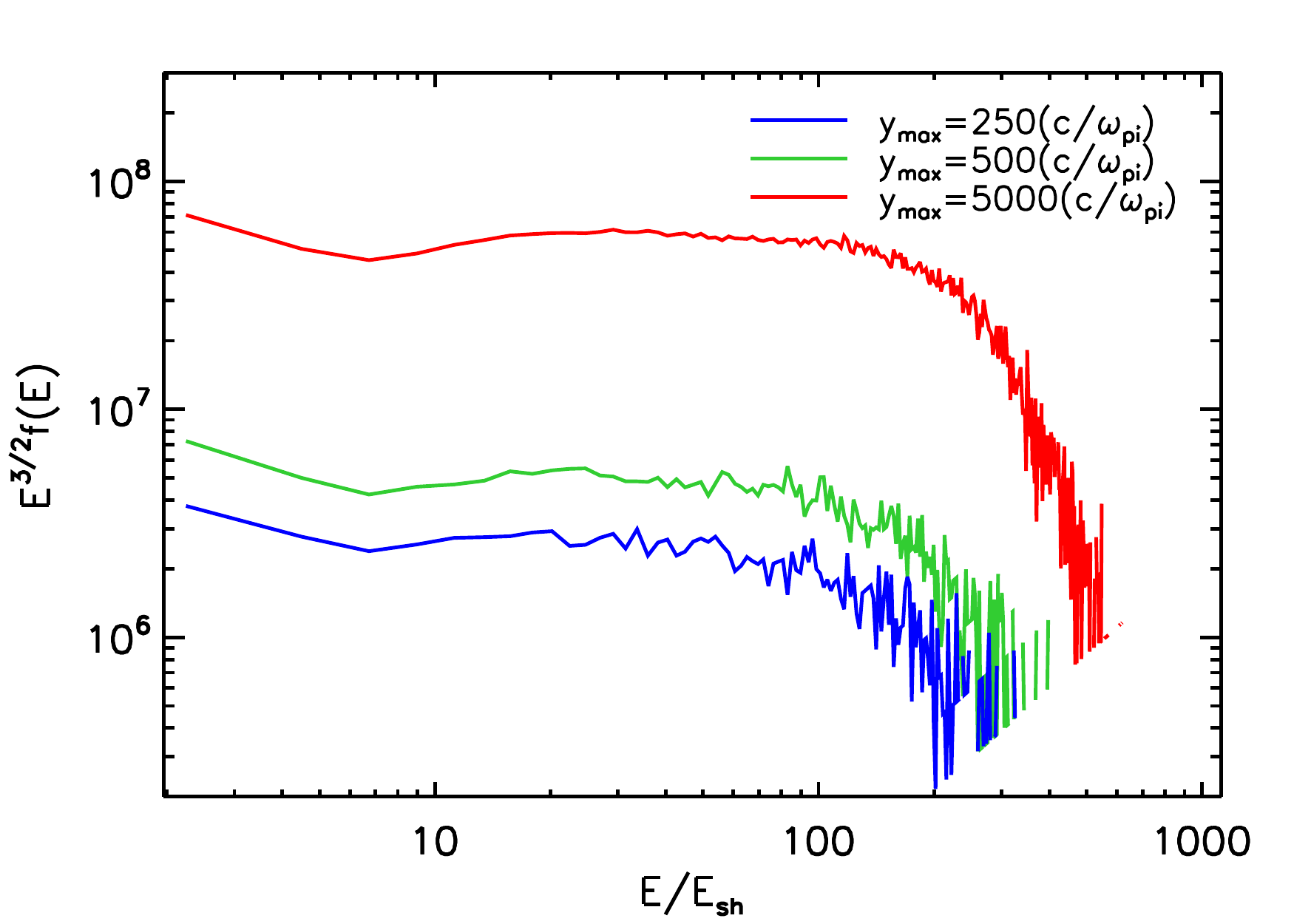}
  \caption{Saturated particle energy spectra for three different transverse domain sizes. Particle energization is seen to increase with the domain size.}
  \label{compare_y}
  \end{figure}

 \begin{figure}
  \centering
     \includegraphics[width=0.48\textwidth,angle=0]{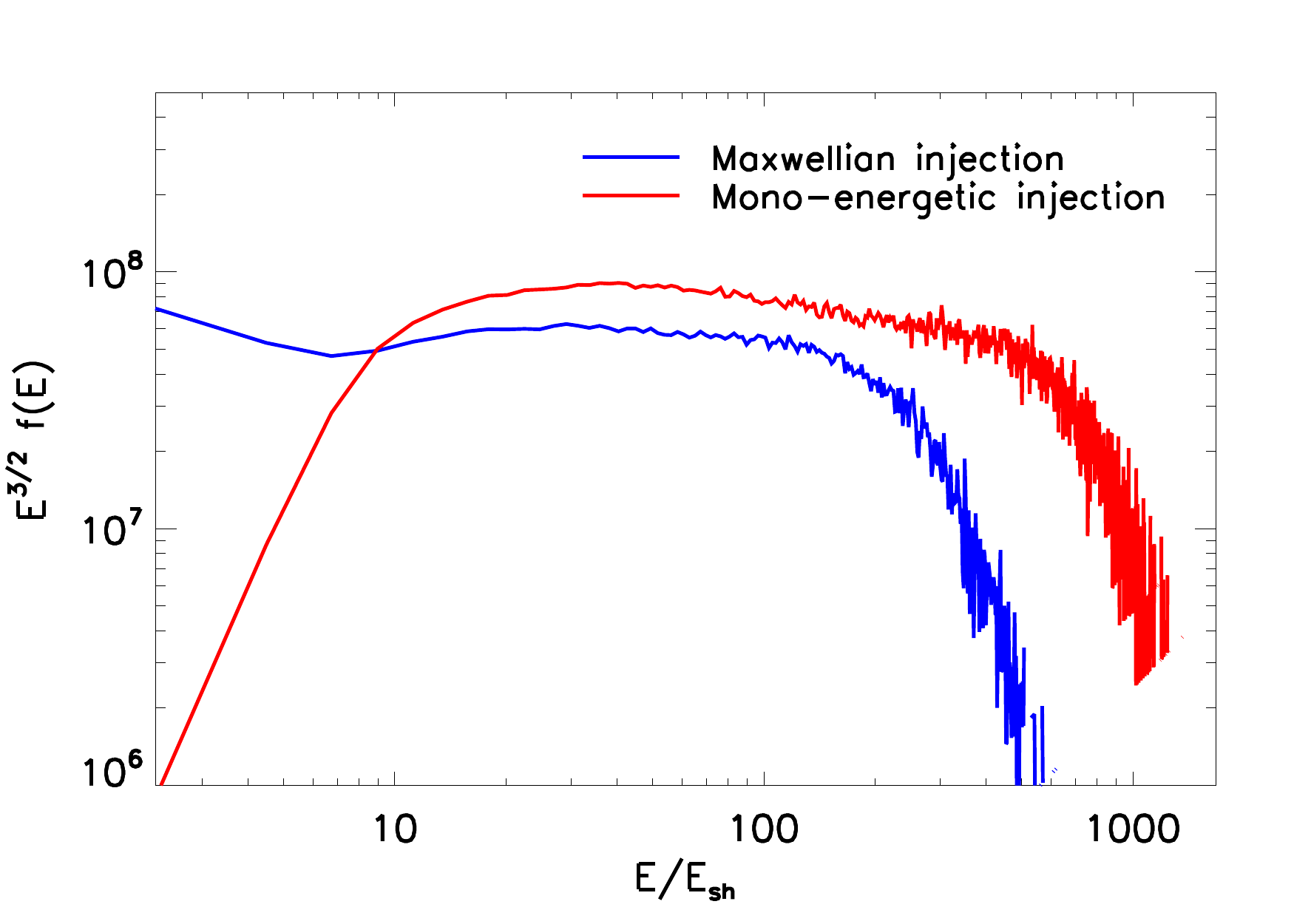}
  \caption{Saturated particle energy spectra for mono-energetic (E$_{inj}$ = 10E$_{sh}$) and maxwellian injection. }
  \label{compare}
  \end{figure}

The maximum energy of the SEPs during CME shocks, however, has been seen to have reached 100 MeV~\citep{Russell_2013}. A possible reason behind not being able to produce such highly energetic protons in our simulation could be the size of the simulation box, which restricts the longer wave modes. Figure~\ref{compare_y} compares the energy spectra for three different simulations with varying transverse box sizes, 250, 500 \& 5000 $(c/\omega_{pi})$, during their near-saturation phase. The tail of protons' energy spectrum extends towards the higher energy with an increase in transverse box size, thereby indicating the role of the same in constraining the maximum energy gained by the protons.\\

In reality, solar wind demonstrates characteristic fluctuations. The ambient solar wind of our present simulation is initiated without any turbulent fluctuation. We envisage that introduction of such turbulent fluctuation in the initial atmosphere may play an important role in the particle energization process.\\

We have also compared our Maxwellian injection recipe with the mono-energetic injection recipe, where $E_{inj} = 10E_{sh}$, typically employed by previous investigators~\citep{Bai_2015, Mignone18}. In section~\ref{sec:injection}, we have already mentioned that in our present simulation, we choose to have a Maxwellian injection with downstream plasma temperature. For this case we had carefully chosen the combination of $\eta =3 \times 10^{-3}$ and the number of particles per cell to be 10, so that the particles do not disrupt the shock. On the other hand, the mono-energetic injection with the above combination of $\eta$ and the number of particles per cell disrupts the shock indicating over-injection. We, therefore, have chosen to bring down $\eta$ and also the number of particles per cell for the mono-energetic injection case to $2 \times 10^{-3}$ and $4$, respectively. A comparison between the two energy spectra in the case of two injections are shown in Figure~\ref{compare}. In the case of Maxwellian injection, particles are seen to have gained energy up to $500 E_{sh}$ (equivalent to 10 MeV), whereas for mono-energetic injection recipe particles have gained energy up to $1000 E_{sh}$ (equivalent to 20 MeV). The possible reason behind this discrepancy is that the particles have wide energy distribution in Maxwellian injection. Lower energy particles in the distribution have low probability of participating in the energization process. On the other hand, in mono-energetic injection, all particles being already suprathermal, the entire population may participate in the acceleration process and thus may end up with a higher energy tail in the energy spectrum.

\subsection{Quasi-perpendicular shock}
\begin{figure}
  \centering
     \includegraphics[width=0.48\textwidth,angle=0]{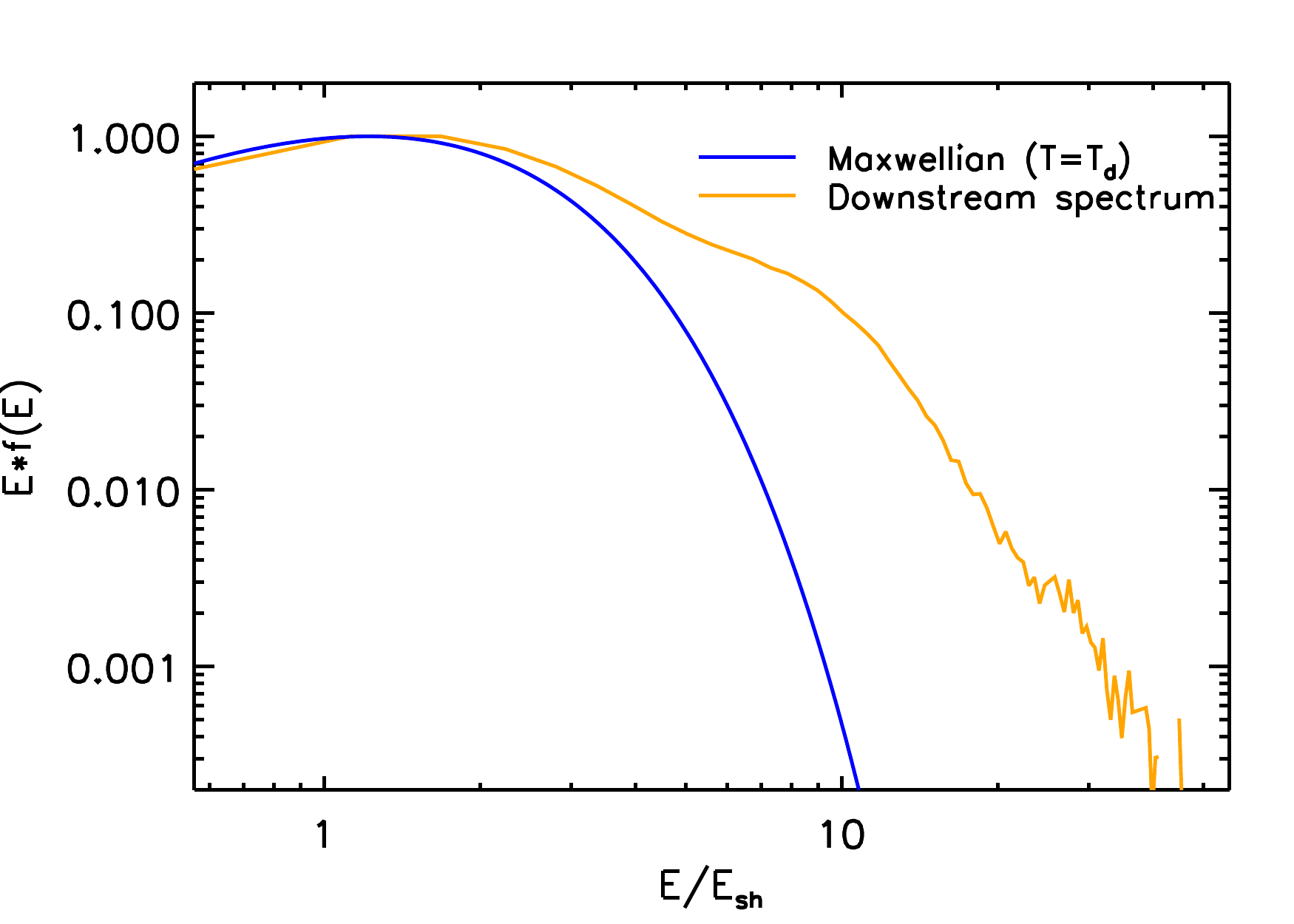}
  \caption{Injected Maxwellian and saturated energy spectra for quasi-perpendicular shock with Mach number $\approx$ 19.}
  \label{1D_spectra_M19_qperp}
  \end{figure}

It is well known that the particle acceleration mechanism in quasi-perpendicular shocks is much different than the parallel shocks. We, therefore, simulate a quasi-perpendicular shock with the same Alfv\'enic Mach number ($\approx 19$). This time, the initial magnetic field makes an angle $\psi=75^{\circ}$ with the shock normal direction, making the shock quasi-perpendicular. The corresponding magnetic field components are now $B_{x} = B_{bg}\cos(\psi)$ and $B_{y} = B_{bg}\sin(\psi)$ with $B_{bg} = 3.55B_{0}$ as mentioned in section \ref{sec:set_up}. Initially particles are injected following the same prescription mentioned in section~\ref{sec:injection}. Figure~\ref{1D_spectra_M19_qperp} shows the particle energy spectrum at the end of the simulation. It shows that the particles only get energized up to $40-45E_{sh}$.   \\

\begin{figure}
  \centering
       \includegraphics[width=0.48\textwidth,angle=0]{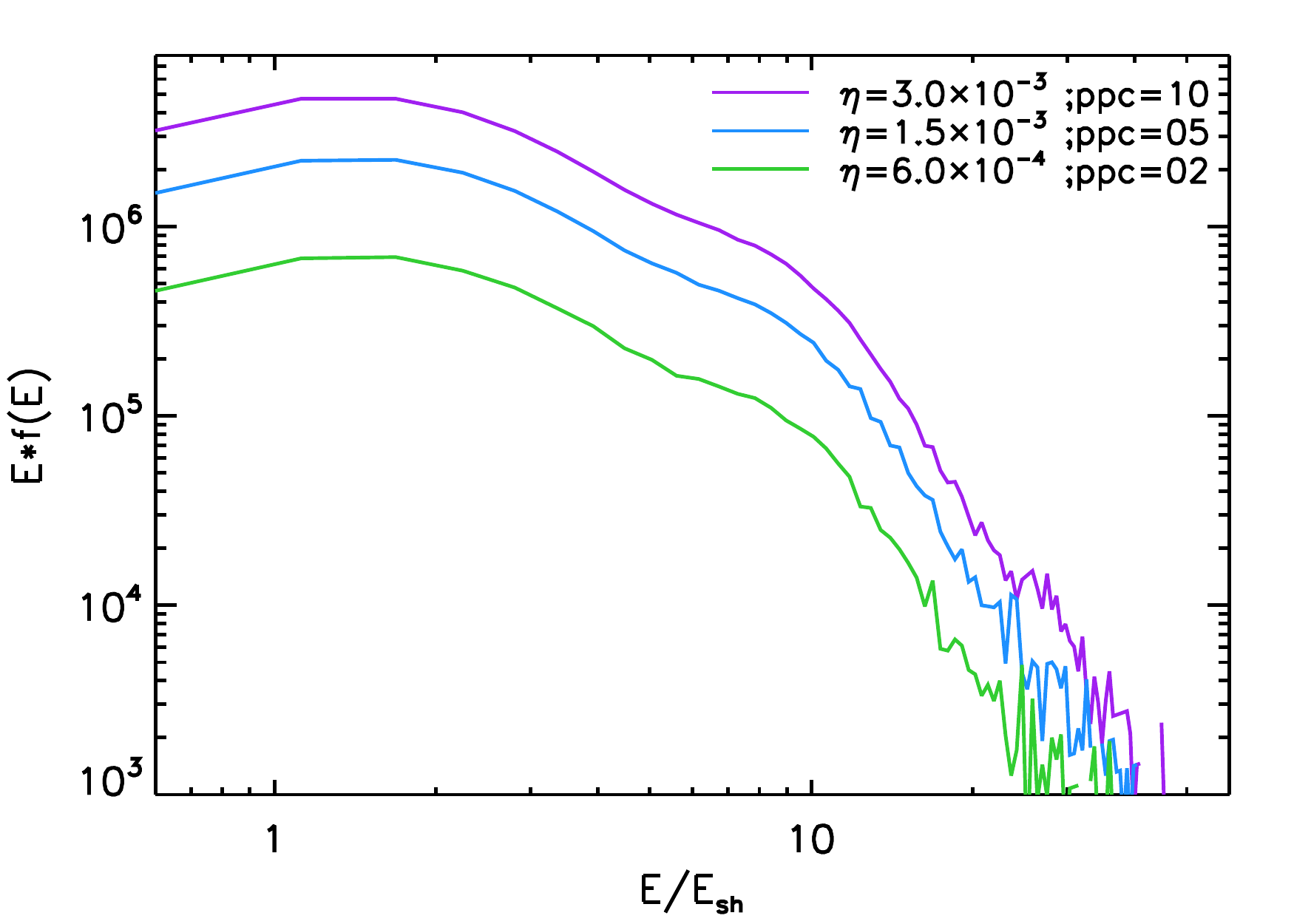}
  \caption{Energy spectra derived from the simulation of quasi-perpendicular shock with $M_{A}\approx19$, by changing injection fraction ($\eta$) and particles per cell (ppc).}
  \label{compare_eta}
  \end{figure}

Earlier hybrid-PIC \citep{Caprioli_2014a} simulation has shown a decrease in acceleration efficiency for quasi-perpendicular shock. Naturally, this suggests the injection of less number of suprathermal particles in the present case. We ran our quasi-perpendicular simulation for two more combinations of $\eta$ and the number of protons per cell, keeping the mass density of individual protons same as our previous injection. The two sets include $\eta = 1.5 \times 10^{-3} $ and $6 \times 10^{-4}$ with $5$ and $2$ protons per cell, respectively. The result is compared with our previous simulation where we considered $\eta = 3\times 10^{-3}$ and 10 protons per cell.
Our new choices ensure less amount of suprathermal particles in the domain. Though the resultant energy spectra from all three combinations vary in terms of the total number of protons, all of them show a very similar energization (Figure~\ref{compare_eta}), indicating the fact that the population of injected suprathermal protons does not affect the overall energization process in quasi-perpendicular shock. The following analyses are performed on the simulation done with the injection prescribed in Section \ref{sec:injection}.\\

\begin{figure}
  \centering
     \includegraphics[width=0.48\textwidth,angle=0]{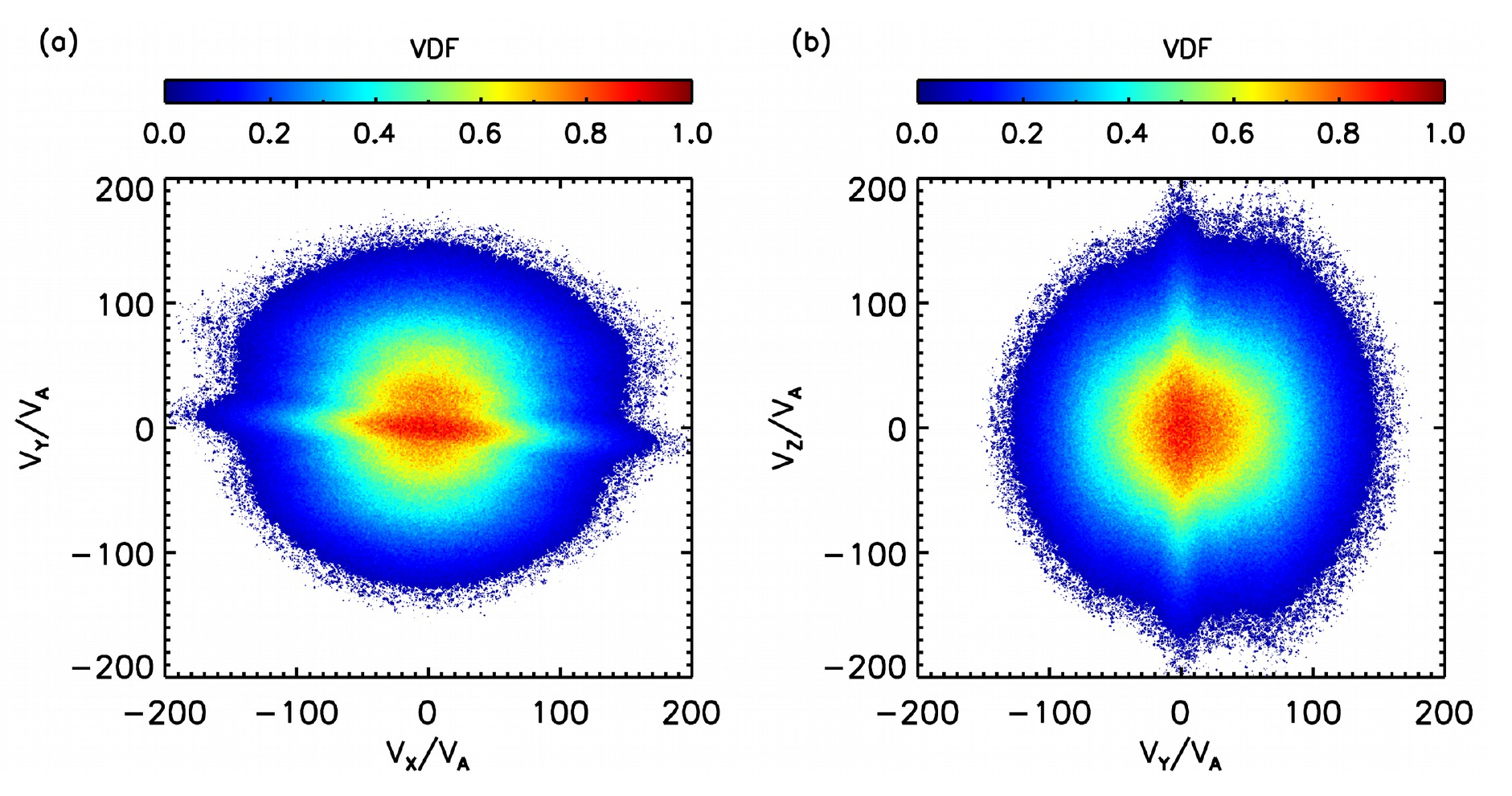}
  \caption{2D velocity distribution function of the quasi-perpendicular shock with $M_A \approx$ 19. Particles were collected from just behind the shock. The region is marked in Figure~\ref{angle_by_bx_bz}}.
  \label{2D_vel_dist}
  \end{figure}

 To understand the energization process, we need to understand which velocity component gains energy. Figure~\ref{2D_vel_dist} depicts 2-D particle velocity distribution of the downstream region. Both the velocity distribution functions show anisotropic nature. Figure~\ref{2D_vel_dist} (a) shows that the velocity distribution $V_x-V_y$ is more skewed along an axis making $\approx 4^{\circ}$ with the $V_x$ axis, while $V_y-V_z$ is more skewed along the $V_z$ axis. This indicates particles are getting accelerated on a plane making an angle $\approx 4^{\circ}$ with the X-Z plane. \\

\begin{figure}
  \centering
      \includegraphics[width=0.48\textwidth,angle=0]{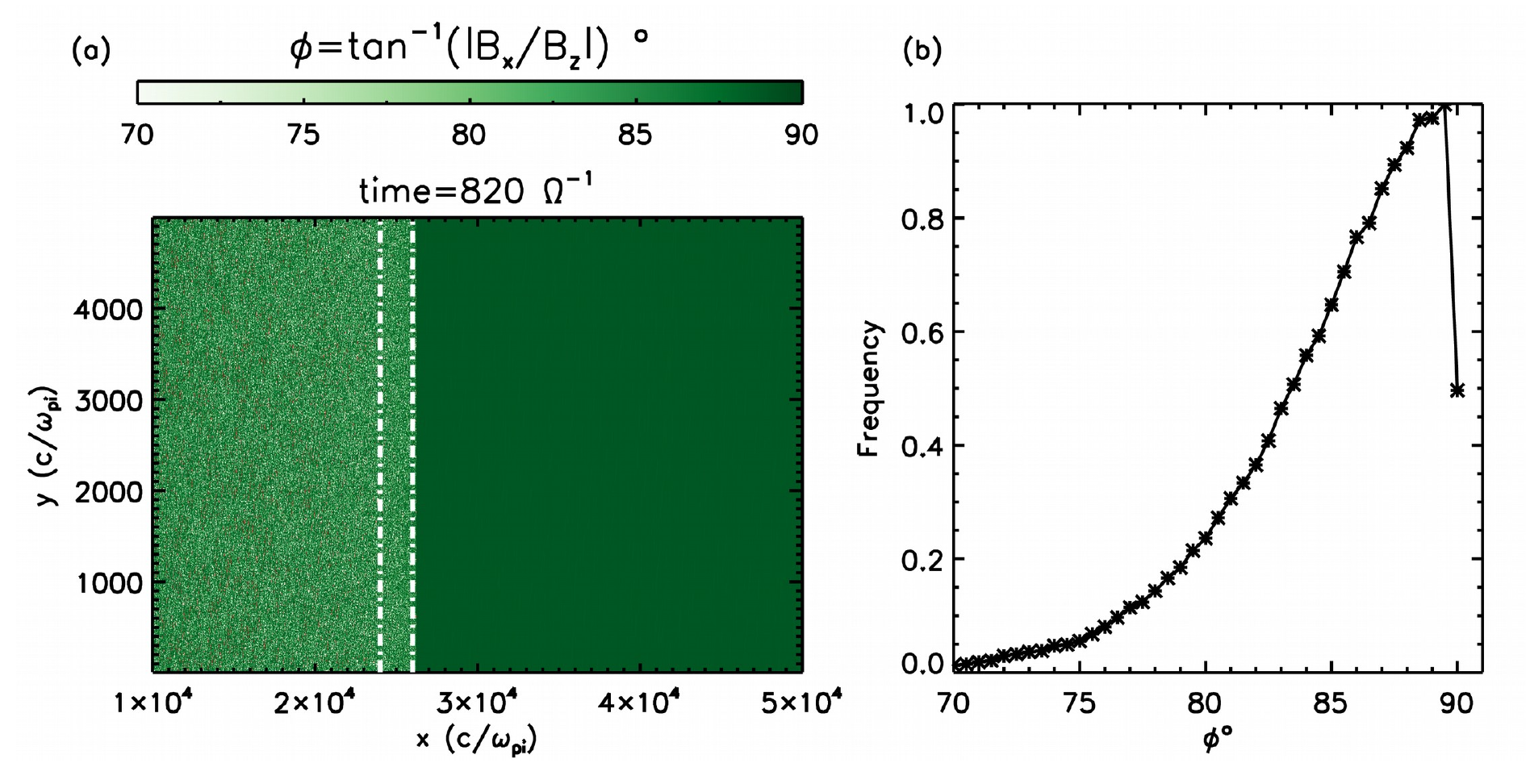}
      \includegraphics[width=0.48\textwidth,angle=0]{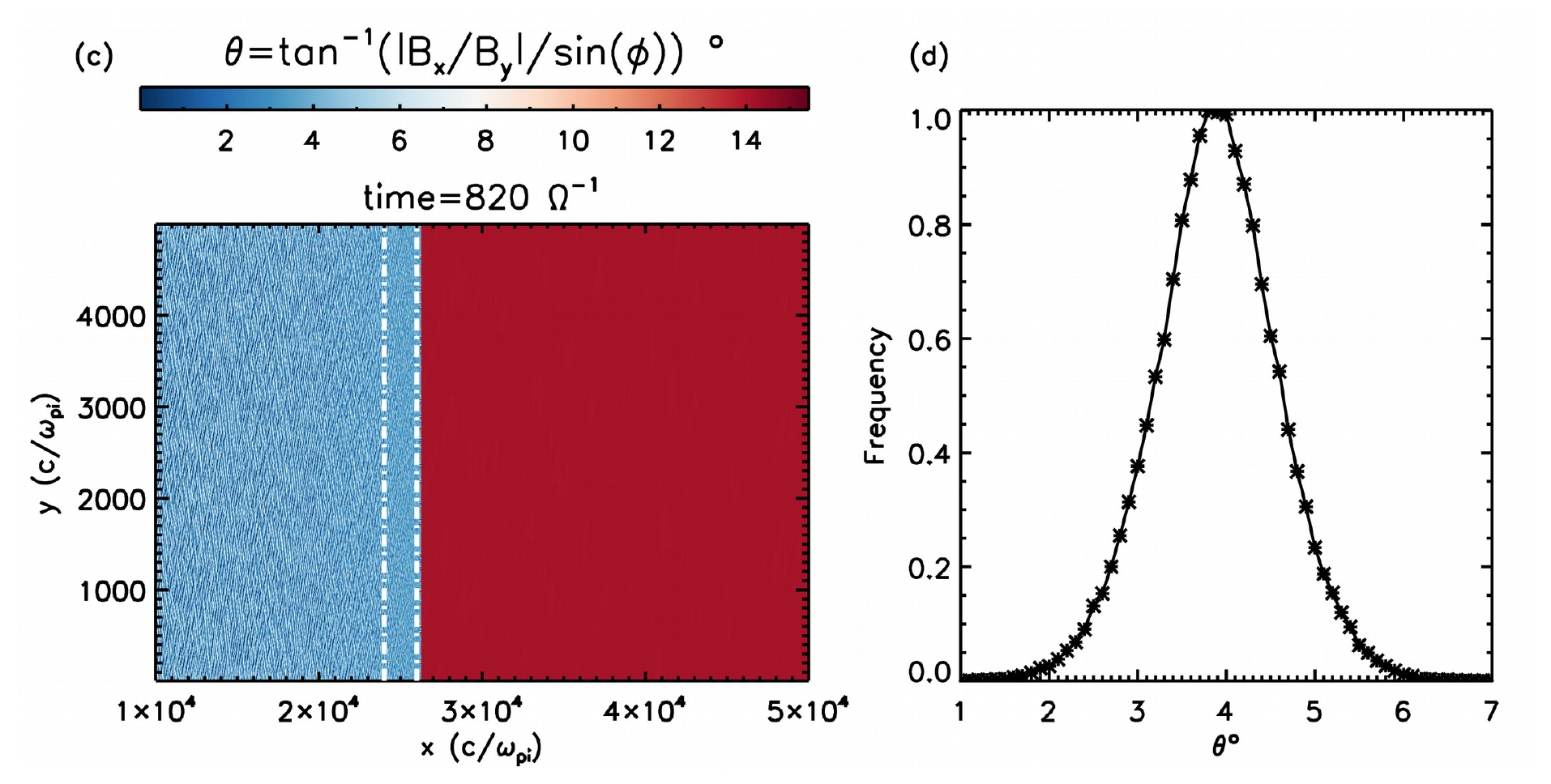}
  \caption{The orientation of the magnetic field just behind the shock. The particular region of interest is marked with vertical dashed lines (a) Contour plot of the angles ($\phi$) between the Y-Z plane and the projection of $\bm{B}$ onto the X-Z plane. (b) A histogram depicting the dominant $\phi$ angle of the region of interest, showing that the mean magnetic field of the region lies on the X-Y plane. (c) Contour plot of the angle between $\bm{B}$ and the Y axis. (d) Histogram plot of the region of interest to show the dominant $\theta$ angle showing the mean field makes an angle $\approx 4^{\circ}$ with the Y axis. Particle velocity distribution function of the same region is plotted in Figure~\ref{2D_vel_dist}}.
  \label{angle_by_bx_bz}
  \end{figure}

 We further investigated the mean magnetic field direction of the downstream region of interest (ROI) during the simulation's saturation phase. Here we remind the reader that this is a 2.5D simulation. Therefore, even though computation is performed only on a 2D plane, all three components of a vector field are evolved in the simulation. As our simulation is performed in the X-Y plane, we assume that the local magnetic field $\bm{B}$ makes an angle $\theta$ with the Y-axis. To accommodate the 3rd component of the magnetic field, we also assume $\phi$ to be the angle between $\bm{B}$'s projection onto the X-Z plane and the Y-Z plane. We derive the value of $\theta$ \& $\phi$ values at every grid of the ROI and the respective histograms are plotted to see the overall orientation of the field. Figures~\ref{angle_by_bx_bz} (a) \& (b) demonstrate that in the ROI $\phi \approx 90^{\circ}$ is the dominant angle, which implies the mean $\bm{B}$ lies on the X-Y plane. Figure~\ref{angle_by_bx_bz} (c) \& (d) demonstrates that in the same downstream region $\theta \approx 4^{\circ}$ is the dominant angle. Overall Figure~\ref{angle_by_bx_bz} concludes that the mean $\bm{B}$ lies in the X-Y plane and makes an angle $\approx 4^{\circ}$ with the Y-axis. \\

After knowing the accelerated components of the particles' velocity and the orientation of the mean magnetic field in the ROI, it is now understood that particles are predominantly getting accelerated in the plane perpendicular to the mean magnetic field (i.e., the plane making an angle $\approx 4^{\circ}$ with the X-Z plane). The convective electric field ($\bm{E}=-\bm{u} \times \bm{B}$) is the reason behind this. The field acts on the plane perpendicular to the mean magnetic field and accelerates protons on the same plane. Concurrently, the local \textit{grad} B force, developed across the shock front due to shock compression, makes the particles drift on the plane of the shock. Additionally, particles' velocity components perpendicular to the mean magnetic field, namely the $V_{\perp}$ component, increases in magnitude. The parallel component $V_{\parallel}$ still maintains the injected distribution, resulting in an anisotropic 2-D velocity distribution behind the shock. The mechanism is popularly known as SDA and plays a major role in accelerating protons in our quasi-perpendicular shock simulation.\\

\begin{figure}
  \centering
     \includegraphics[width=0.48\textwidth,angle=0]{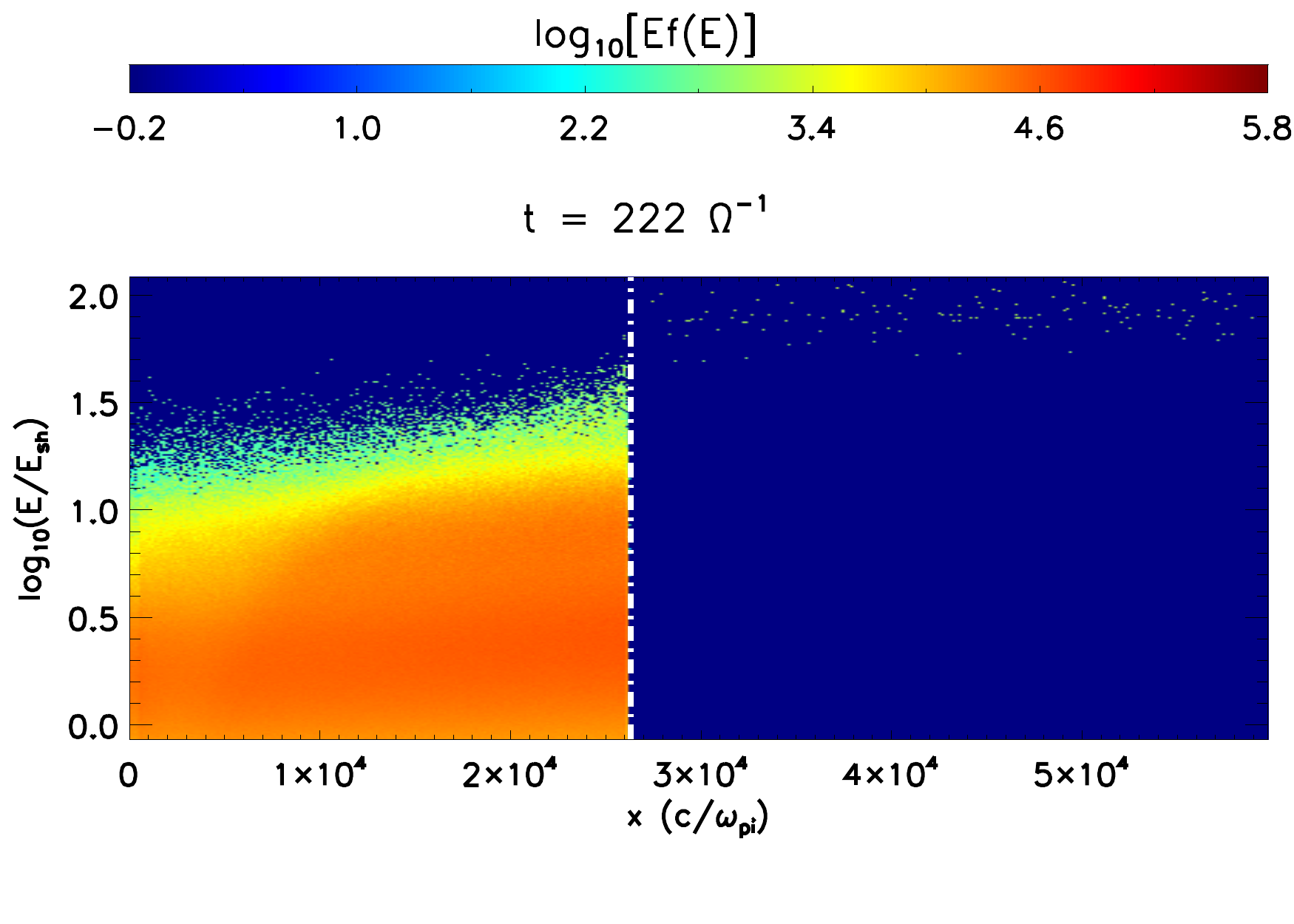}
  \caption{2D energy spectrum, showing the particle energy distribution as a function of position $x$. Here $E_{sh}$ is the shock energy given by $E_{sh} = v_{u}^2/2$ where $v_u$ is the upstream plasma speed in the downstream reference frame. Location of the shock front is indicated by the white dashed line. As evident from this plot, high energy particles outrunning the shock are very less in number and hence cannot seed strong fluctuations in the upstream.}
  \label{2D_spectra_per}
  \end{figure}

We should point out that although the Alfv\'enic Mach of both the parallel and quasi-perpendicular shocks are same in our above mentioned simulations, the energy gained by the protons in the quasi-perpendicular shock case is significantly less. This can be explained with Figure~\ref{2D_spectra_per}. Only a few high energy protons are seen to have escaped the shock front and move into the upstream region. After getting accelerated by SDA, protons are mostly advected into the downstream region within a few gyrocycles, resulting in the truncation of further energization process. The few high energy protons that managed to escape the shock front are insufficient to generate strong fluctuations in the upstream region. Nearly non-fluctuating density ($\delta\rho/\rho_{0}\approx0$) in the upstream implies that there is no active participation of DSA in our simulation to accelerate protons. However, in reality, inherent fluctuations of the SW can excite DSA and may accelerate particles further. This will be investigated in our future work.\\

 \section{ICME simulation for low mach shock}\label{sec:lowmach}
ICME-shocks with low Alfv\'enic Mach numbers \citep{Berdichevsky_2000, Oh_2007,Lugaz_15} are more common. Keeping that in mind, we also have simulated ICME-shocks with low  Alfv\'enic Mach $\approx 10$ and $4$. Generation of shock and particle injection recipes are identical to the cases discussed in section \ref{sec:mach_19}. The only parameter that is changed in order to produce shocks with low Machs is the relative velocity between the SW and ICME plasma. For Alfv\'enic Mach $\approx 10$, we have injected a SW flow with relative speed $-50.05\,v_{A}$ $(= - 976$ km s$^{-1})$, whereas for Alfv\'enic Mach $\approx 4 $ the relative speed is $-20\,v_{A}$ $(= - 390$ km s$^{-1})$. Both these shocks are parallel in nature. \\

\begin{figure}
  \centering
     \includegraphics[width=0.48\textwidth,angle=0]{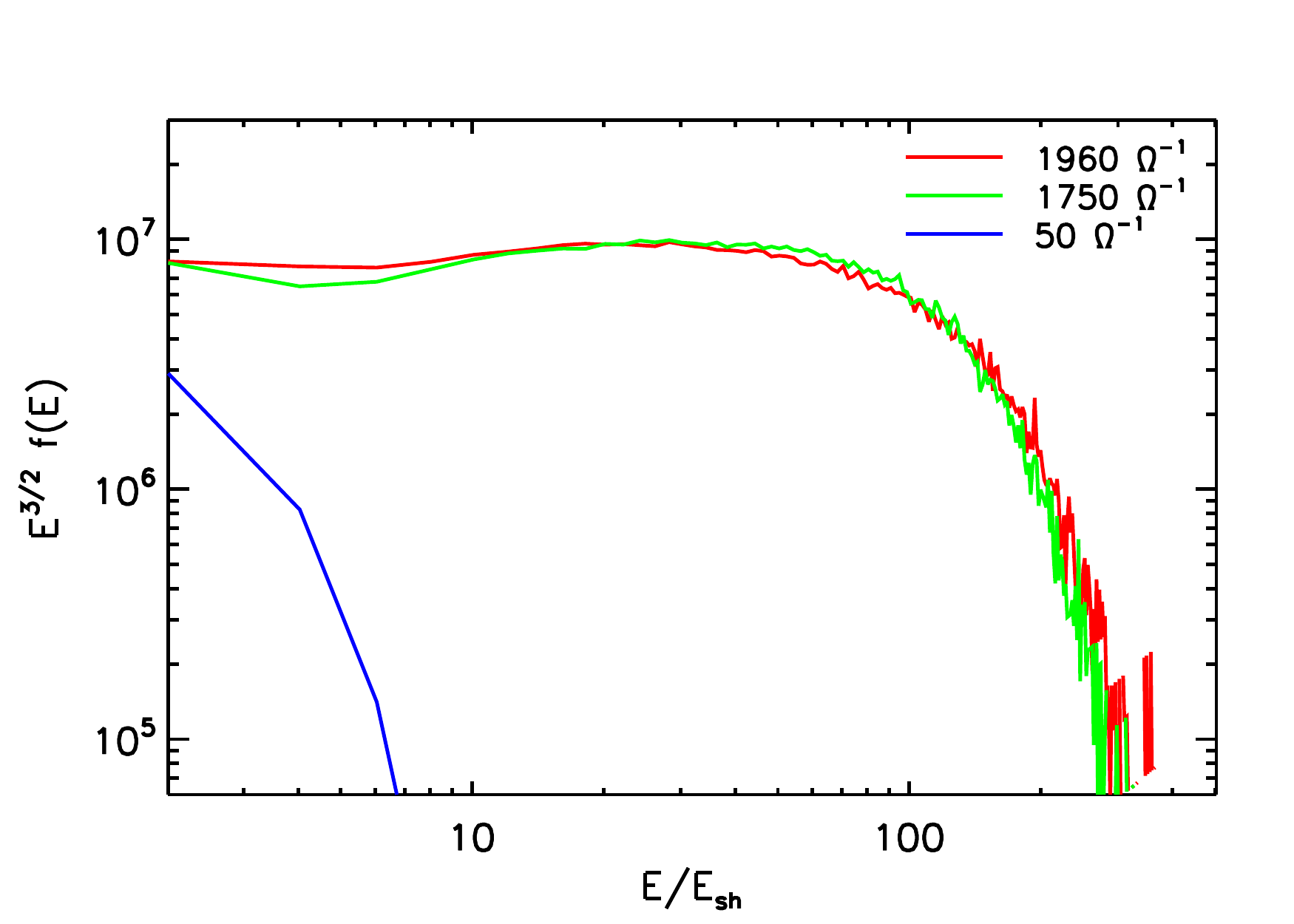}
  \caption{Evolution of downstream spectra for a parallel shock with $M_{A}$ $\approx 10$, calculated over a region of width $1000(c/\omega_{pi})$ just behind the shock.}
  \label{1D_spectra_M10}
  \end{figure}

\begin{figure}
  \centering
     \includegraphics[width=0.48\textwidth,angle=0]{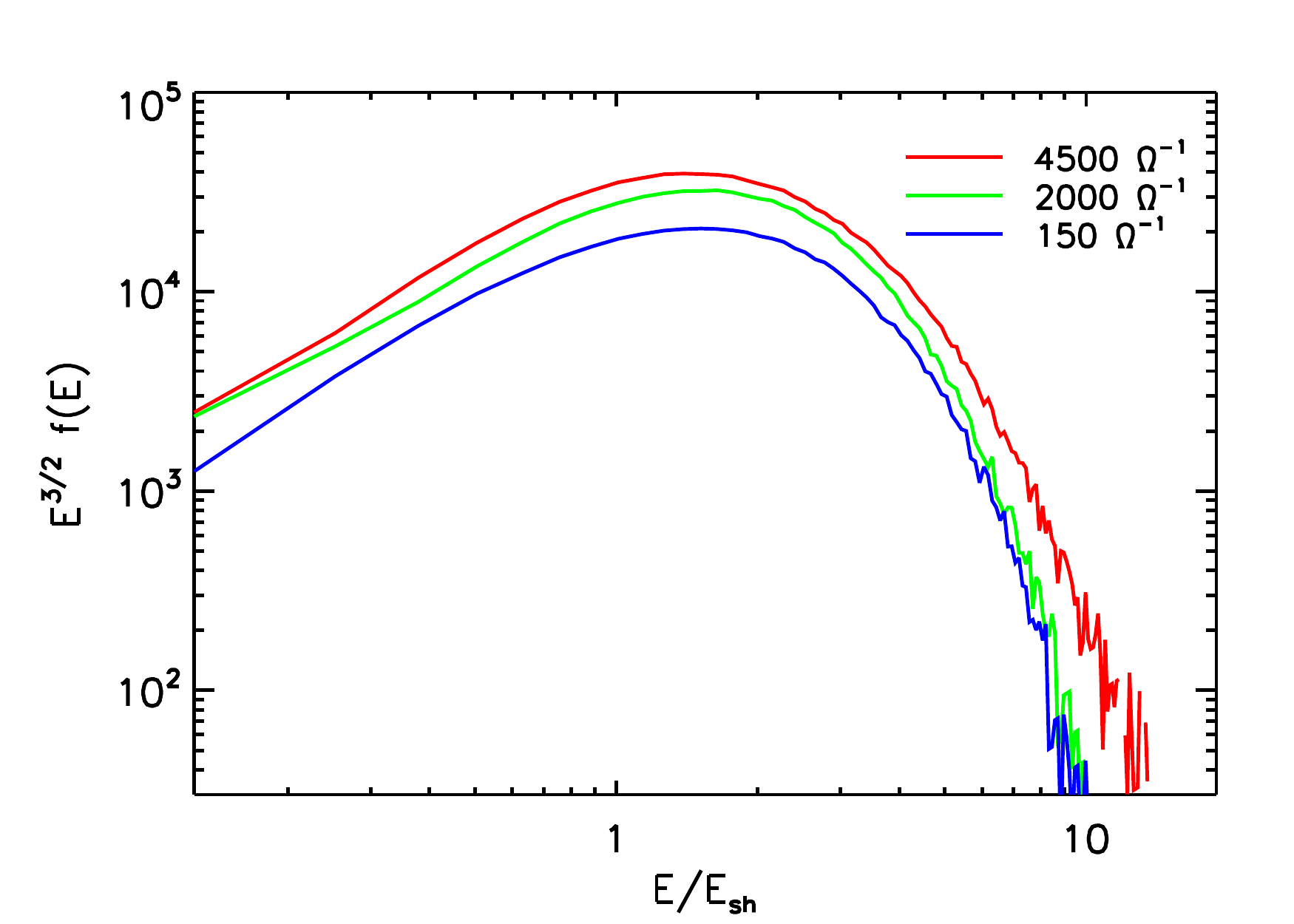}
  \caption{Evolution of downstream spectra for a parallel shock with $M_{A}$ $\approx 4$, calculated over a region of width $1000(c/\omega_{pi})$ just behind the shock.}
  \label{1D_spectra_M4}
  \end{figure}

Figure~\ref{1D_spectra_M10} exhibits the particle energy spectrum for the $M_A \approx 10$ shock. Particles are seen to get energised up to $300 E_{sh}$ in this simulation. The flat part of the spectrum directs towards the DSA mechanism for particle acceleration. On the other hand, Figure~\ref{1D_spectra_M4}, which depicts particle energy spectrum for the shock with Alfv\'enic Mach $\approx 4$, shows no indication of particle acceleration. \\

Therefore, these sets of numerical simulations follow our intuitive guess that Alfv\'enic Mach number is a good indicator of shocks' capability to accelerate particles. Low Mach shocks may not work as good accelerators.\\

\section{Summary and discussion}\label{sec:summary}

 This paper has investigated the particle acceleration process in  ICME shocks. We used the MHD-PIC module of the PLUTO code \citep{Mignone18}. The neutral background plasma is treated as a magnetohydrodynamic fluid, and non-thermal proton dynamics is taken care of using a PIC code. The reactions of particle dynamics onto the MHD fluid and vice versa are appropriately managed. Compared to earlier heliospheric shock models \citep{Zank_2000}, this model is more local in nature. Unlike the former, it can only simulate a small region across the shock. But, particles evolve self consistently on the background thermal plasma.  This hybrid model is more similar to \cite{Gargate_2014}'s model, but computationally much less expensive. Such a hybrid simulation enables us to follow the particle dynamics and fluid evolution until the particle energy spectrum nearly saturates. These simulations empower us to predict various \textit{in situ} observations and help us plan future missions. \\

We started our investigation with a CME shock having an extremely high Alfv\'enic Mach number ($M_A \approx 19$). Even though such a strong shock is rare in the heliosphere, one was evidenced lately (on 2012 July 23). The particle acceleration process has been probed initially for a parallel shock, where the shock propagation direction exactly matches with the ambient solar wind magnetic field. This being an idealistic case, we also have studied a quasi-perpendicular shock, where the ambient magnetic field forms a $75^{\circ}$ angle with the propagation direction. CME shocks with lower Mach numbers (e.g., $M_A \approx$ 10 \& 4)  have also been investigated in the present study. Particles are injected at the shock front in all our simulations. During injection, particles' distribution follows a Maxwellian with characteristic downstream temperature (i.e., energy $E_{sh}$). In the following, we outline our results as well as suggest ways to further explore the observation data that may confirm various physical processes occuring around such CME shocks.\\

\textit{Instability and dynamo in the parallel shock} -- The first simulation, which delineates a plane parallel shock with high Alfv\'enic Mach ($M_A \approx 19$), demonstrates that such a shock can energize protons up to $500 E_{sh} (\approx 10 MeV)$. The slope of the particle energy spectrum (Figure~\ref{1D_spectra_par}) indicates the role of DSA in the particle acceleration process. We also note that once particles get energized, many of them leave the shock front and move to the far upstream region (Figure~\ref{2D_spectra}). \\

The upstream of the shock is seen to become turbulent with the growth of density cavities. Such cavities are due to the development of non-resonant Bell instability in the upstream region. As the instability grows, the sizes of these cavities increase. Once the local ion-gyroradius becomes comparable to the cavity size, the instability ceases to grow (Figure~\ref{power_rho_norm}). We have further investigated the existence of resonant streaming instability in the domain. We find that the longest and shortest resonant modes encompass the peak of the magnetic power spectrum. However, the fastest growing mode of non-resonant Bell-instability turns out to be responsible for the peak of the power spectrum of the region's transverse magnetic field (Figure~\ref{bell}).  \\

The turbulent magnetic field of the shock downstream region is seen to have enhanced more than is expected from the shock compression, hinting at the presence of a dynamo mechanism in the region. Due to non-aligned pressure and density gradient at the corrugated shock front, instability like Richtmyer-Meshkov instability may kick in and develop turbulence. Such turbulence stretches the field lines and thereby enhances the local magnetic field. Kinetic or magnetic diffusivity are not explicitly specified in our simulation. Nevertheless, when we compare the kinetic and magnetic energy spectra (Figure~\ref{v_b}) the dominant magnetic energy towards the small scale ($k > 0.07~(c/\omega_{pi})^{-1}$) indicates the development of small scale dynamo in the region. \\

\textit{Quasi-perpendicular shock demonstrates velocity anisotropy} -- 
We have also performed simulation of quasi-perpendicular shock with a high Alfv\'enic Mach number ($M_A \approx 19$). The particle energy spectrum shown in  Figure~\ref{1D_spectra_M19_qperp} indicates energization up to $40-45E_{sh}$ which is much smaller than the particle energisation in the parallel shock with the same Alfv\'enic Mach. In quasi-perpendicular shock, the particle energization is expected due to Shock Drift Acceleration process (SDA), where the convective electric field $\bm{E}=-\bm{u} \times \bm{B}$ plays a vital role in accelerating the particles. Activation of SDA in the present case is evident from the particles' velocity distribution as shown in Figure~\ref{2D_vel_dist}. The distribution shows the development of anisotropy on the plane perpendicular to the local mean magnetic field.

Our simulations demonstrate that, in quasi-perpendicular shocks (Figure~\ref{2D_spectra_per}), high energetic particles rarely escape to the upstream and therefore do not develop strong turbulence in the region, as it happens in the parallel shock. This also prevents the particles from participating in DSA and accelerating further. However, once again, the SW turbulence may play a crucial role in accelerating these particles, which is not incorporated in the present simulations.  \\

\textit{Low Mach shocks are less effective to accelerate particles} -- As expected, our simulations of low-Mach number shocks show no effective particle acceleration. We have simulated parallel shocks with Alfv\'enic Mach, $M_A \approx$ 10 and 4. Though the shock with $M_A \approx$ 10 could energize particles upto $300 E_{sh}$, the shock with $M_A \approx$ 4 does not show evidence of particle acceleration. \\

It is worth mentioning that our simulations were unable to produce particles with very high energy (e.g., up to 100 MeV, as in the case of real observations~\cite{Russel_13}). Multiple reasons can be responsible for this. While traveling from the Sun to 1 AU, the CME shocks encounter variable backgrounds over a long duration of time. This may help particles to get energized continuously. On the other hand, our localized simulations are provided with a background atmosphere of 1 AU, and therefore can only energize particles to whatever energies is possible over this background.  Furthermore, our present simulations start with a uniform solar wind background. The inherent fluctuations of the real solar wind may energize particles to a greater extent, which needs to be investigated in the future. The restricted input energy distribution of the particles may also be responsible for the limited energization process. In the simulations, we inject particles with a Maxwellian distribution having a characteristic temperature of the downstream. However, these may not be the only particles getting induced at the shock front in reality. Some amount of much higher energetic particles that are already energized due to magnetic reconnection may also get induced at the shock front for further processing and may give rise to the observed high energy tails. Such a possibility has been proposed earlier~\citep{Klein_2001}. As others (e.g., \cite{Bai_2015, Mignone18}), we also have attempted to inject particles with $10 E_{sh}$. The resultant energy distribution gets extended towards the higher end but only by a few MeVs as compared to the Maxwellian injection (Figure~\ref{compare}). We have also shown that the limited size of the simulation domain may restrict the growth of more extended modes in the domain and regulate energy growth beyond a threshold. \\

Understanding particle acceleration in shocks is a major topic in overall astrophysics. Shocks of different strengths and magnetic configurations are ubiquitous in the heliosphere. Heliospheric shocks can be quasi-parallel or quasi-perpendicular, depending on the relative angle between their propagation direction and ambient upstream magnetic field. Therefore, shocks that are potential accelerators (relatively high $M_A$) should be able to demonstrate the characteristics of both DSA and SDA mechanisms. The issue of particle acceleration, therefore, can be explored directly by using heliospheric \textit{in situ} data. In the following paragraph, we highlight a few observables that can confirm our current understanding of the subject.\\

In principle, SEP energy spectra should be able to demonstrate the characteristic -3/2 slope if DSA is the dominant mechanism. On the other hand, the 2D velocity distribution of the particles should be able to demonstrate anisotropy if SDA plays a pivotal role in accelerating particles. The existence of non-resonant Bell instability can be verified by measuring the size of the upstream density cavities or by analyzing the upstream magnetic turbulence. It will also be interesting to see if the downstream magnetic field gets enhanced due to turbulent dynamo. \\

 It is possible to perform local numerical simulations of wide ranges of ICME shocks and compare the simulation outputs with real observations. ICME shocks with varied strength at different heliospheric distances  can be compared with the simulation results. We can also corroborate the effect of different solar wind background plasma on ICME shocks.  We envisage data from the Integrated Science Investigation of the Sun (IS\(\odot\)IS) and Electromagnetic Fields Investigation (FIELD) experiments on board the Parker Solar Probe (PSP), the Energetic Particle Detector (EPD), Solar Wind Plasma Analyser (SWA) and Magnetometer (MAG) on board Solar Orbiter or the Aditya Solarwind Particle EXperiment (ASPEX) on board Indian Space Research Organisation’s  Aditya-L1 mission will be able to probe in this direction. \\

\begin{acknowledgments}
Computations were carried out on the Physical Research Laboratory's VIKRAM cluster. We thank the anonymous referee for her/his insightful comments, which helped us improve the manuscript. BV would like to thank the support provided by the Max Planck Partner Group established at IIT Indore.
\end{acknowledgments}

\bibliography{sample}{}
\bibliographystyle{aasjournal}

\end{document}